# Disentangling Single- and Biexciton Dynamics with Photoelectron-Detected Two-Dimensional Electronic Spectroscopy

*Luisa Brenneis[1], Matthias Hensen[1], Julian Lüttig[1,2,*], Tobias Brixner[1,3,*]*

[1]Institut für Physikalische und Theoretische Chemie, Universität Würzburg, Am Hubland, 97074 Würzburg, Germany

[2]Department of Physics, University of Ottawa, 150 Louis-Pasteur Pvt, Church St, Ontario, ON K1N 6N5, Canada

[3]Center for Nanosystems Chemistry (CNC), Universität Würzburg, Theodor-Boveri-Weg, 97074 Würzburg, Germany

Corresponding Authors

*E-mail: julian.luettig@uni-wuerzburg.de

*E-mail: tobias.brixner@uni-wuerzburg.de

Action-detected two-dimensional (2D) spectroscopy resolves the time-dependent nonlinear optical response of a quantum system by recording incoherently detected observables such as fluorescence, photoelectrons, or photocurrents which reflect the system's excited-state population. Processes such as exciton–exciton annihilation alter this population and obscure, for instance, energy transfer processes. This limits the information available from action-detected 2D spectra compared to their coherently detected counterparts. Here we investigate time gating and kinetic-energy filtering in photoelectron-detected 2D spectroscopy to disentangle various processes. We implement a numerical simulation protocol that allows us to calculate photoelectron-detected 2D spectra for various systems, demonstrating that time gating can extract the same information as coherently detected 2D spectroscopy, even when annihilation is present. Furthermore, we can directly infer annihilation dynamics. Kinetic-energy filtering additionally enables the isolation of specific excited-state dynamics. Our simulations demonstrate that time gating and kinetic-energy filtering are promising extensions for photoelectron-detected 2D spectroscopy.

**TOC GRAPHICS**

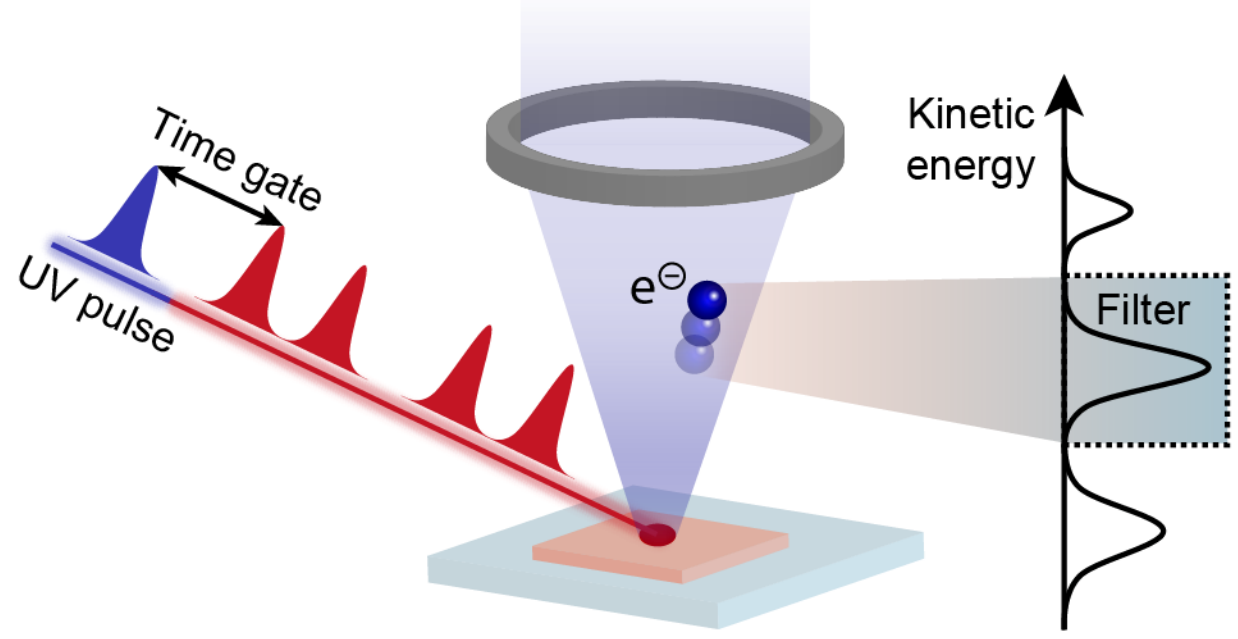

Over the last two decades, several action-detected implementations of two-dimensional electronic spectroscopy (2DES) have been developed. In contrast to coherently detected two-dimensional electronic spectroscopy (C-2DES), action-detected 2DES uses incoherent observables which are usually linked to the population of the excited state. Examples include the detection of fluorescence,[1–6] photoelectrons,[7–10] photoions,[10,11] or internal photocurrent.[12–14] Action-detected 2DES offers several advantages like a sensitivity down to the single-molecule level using fluorescence detection,[15] spatially resolved experiments below the optical diffraction limit by imaging photoelectrons,[8,16] high-frequency-resolution two-dimensional (2D) spectra in the gas phase,[17] and the ability to directly study actual devices, such as solar cells via photocurrent detection.[13] Side-by-side comparisons of coherently and action-detected spectroscopy techniques show similarities but also profound differences, both technical as well as fundamental.[6,18–20]

Action-detected 2DES has several challenges not present in C-2DES. Processes that vary the excited-state population after the pulse sequence has interacted with the system can alter the signal and complicate the interpretation of 2D spectra. Such processes include exciton–exciton annihilation (EEA) or Auger recombination.[21–23] One of the consequences of these processes is the emergence of cross peaks for early population times in the 2D spectra of dimers. In contrast to C-2DES, these cross-peak signals are not exclusively attributable to the exciton delocalization but can also be caused by EEA.[19,22,24,25]

A further challenge in action-detected 2DES of extended systems is incoherent mixing,[23,26,27] i.e., mixing of independent population signals due to EEA. Incoherent mixing can weaken the signatures of single-exciton dynamics in multichromophoric systems. In C-2DES, single-exciton energy transfer is visible as a rise of a cross peak and a decay of the corresponding diagonal peak over the population time $T$. However, in action-detected 2DES, the decay and rise of the corresponding signals occur on a large static background caused by incoherent mixing which makes the signatures of energy transfer difficult to detect.[24] For example, in the light-harvesting 2 (LH2) complex, the rise and decay is found to be only a few percent of the overall signal amplitude of the corresponding peaks using fluorescence detection.[6] In C-2DES of the same system, no static background is present, i.e., the cross peak is absent at early times and emerges basically from zero with the energy-transfer rate. Recently, several strategies to reduce this undesired background have been reported, exploiting spectro-temporal symmetry[28] or polarization-controlled experiments.[26]

The difference between C-2DES and action-detected 2DES is not limited to signatures of coupling and energy transfer. While higher-order C-2DES can be utilized to extract EEA dynamics via the fifth-order signal,[29,30] the sixth-order signal of fluorescence-detected two-dimensional electronic spectroscopy (F-2DES) reflects not only EEA but also overlapping single-exciton contributions such as energy transfer, i.e., it does not exclusively report on EEA. Disentangling such single- and multi-particle contributions in F-2DES requires advanced subtraction schemes.[19]

The question arises as to whether action-detected 2DES can be extended to solve these fundamental issues. Time gating was proposed as a method to recover the same information from fluorescence-detected 2D spectra as from C-2DES despite processes such as EEA.[22] In fluorescence-detected spectroscopy methods, time gating can be implemented by fluorescence upconversion, i.e., overlapping the fluorescence signal with an additional pulse in a nonlinear crystal.[31–34] An alternative strategy to investigate single-exciton dynamics in an action-detected technique is 2D fluorescence excitation (FLEX) spectroscopy, which has been theoretically investigated,[35,36] subsequently explored through proposed experimental implementations and first proof-of-principle studies,[37] and only recently realized experimentally.[38] This spectroscopic technique suppresses certain contributions using fluorescence upconversion, thereby also removing the static background present in F-2DES.

Time gating in action-detected 2DES can also be realized with photoelectron detection. This considerably broadens the range of accessible samples due to the different selection rules associated with the detected observable. In particular, it enables the investigation of systems with low fluorescence quantum yields, such as H-type coupled aggregates, as well as surface systems, where fluorescence emission is quenched. In F-2DES, the system interacts with four laser pulses before the signal is spontaneously generated, i.e., a fluorescence photon is emitted. The fluorescence lifetime is typically much longer than the relaxation dynamics within the single- and double-excited-state manifolds. When applying femtosecond time gating to a fourth-order fluorescence signal, the detection window of each time-gated step captures only a very small fraction of the total fluorescence emission, which is typically spread over nanoseconds. Consequently, the collected signal is extremely weak. In contrast to fluorescence detection, photoelectron detection does not rely on a spontaneous process but on the stimulated process of photoionization. Therefore, the population of system states is probed directly at the moment of photoionization. In some previous photoelectron-detected 2DES (P-2DES) experiments, photoelectrons were generated by the last pulse of a pulse sequence, where each pulse has an identical central frequency.[9,16,39] This substantially decreases the number of accessible signal contributions. Alternatively, an ionization-assisted approach can be envisioned using a fifth pulse which maps the fourth-order response onto a photoemission yield, i.e., the yield of emitted photoelectrons. As a consequence, the detected signal results from six interactions with electric fields: four times through the excitation pulse sequence and twice through the ionization pulse. An ionization-assisted approach was already pointed out and discussed in a previous study[16] and experimentally realized.[7] Such an additional time-delayed ionization pulse inherently introduces a gate time. Implementing an ionization pulse with precise timing relative to the multi-pulse sequence and providing photon energies high enough to induce photoionization adds significant complexity to the setup. However, several recent studies demonstrated the generation of spectrally tunable ultraviolet pulses making the use of an additional ionization pulse feasible.[8,40,41]

Photoelectron detection experiments additionally open the opportunity to perform kinetic energy filtering, e.g., by using a hemispherical energy filter[42] or by a time-of-flight measurement.[43,44] The influence of kinetic-energy filtering in P-2DES was analyzed by Uhl and co-workers for an atomic system.[7] However, this method has not yet been investigated thoroughly for an excitonic system.

Here, we study whether and how the above-mentioned problems, such as hidden coupling features, weak energy transfer signatures, and challenging extraction of EEA rates, can be solved by time gating in P-2DES. We emphasize the ability of the method to disentangle single- and biexciton dynamics within a single experimental approach. To this end, we have expanded our recently published open-source Matlab software package for quantum dynamics simulations, the Quantum Dynamics Toolbox (QDT),[45] to include photoelectron detection as a signal channel. This allows us to model an excitonic dimer system. We use kinetic-energy-resolved P-2DES to track annihilation dynamics by isolating the nonlinear response of specific states. Our simulations show that the two presented concepts—time gating and kinetic-energy filtering—are useful extensions of P-2DES, paving the way for future experiments.

In C-2DES, a signal proportional to the third-order nonlinear polarization is measured.[46] Three pulses, separated by the interpulse delays $\tau$ and $T$, interact with the sample, and the resulting coherent nonlinear signal is detected, for example, through interference with a local oscillator (LO) (Figure 1a, left). The detection axis is thus directly provided, e.g., by a spectrometer, while the excitation axis is obtained by Fourier transformation over $\tau$. In contrast, in action-detected 2DES the interaction of the system with four pulses, separated by the time delays $\tau$, $T$, and $t$, gives rise to the fourth-order nonlinear system response. The separation of the lower-order response and the extraction of the fully absorptive signal contribution, as well as the rephasing and nonrephasing contributions, is possible via phase cycling of the four-pulse sequence.[20] The incoherently detected observable, proportional to the leading fourth-order nonlinear system response, can be generated either spontaneously, e.g., via fluorescence photon emission (Figure 1a, middle) or through a stimulated process, e.g., photoelectron emission (Figure 1a, right), depending on the specific detection channel.

Figure 1b illustrates the principle of P-2DES with an additional ionization pulse. The system is excited by a phase-cycled four-pulse sequence (Figure 1b, red pulses). Subsequently a delayed ionization pulse with a carrier frequency that is higher than that of the other pulses, e.g., an ultraviolet pulse, converts the system's response into the photoelectron-detected signal. We note that the ionizing pulse is not phase-locked to the preceding pulses, as it serves to project the population of the system onto corresponding photoelectron states. The photoelectron yield is recorded as a function of the interpulse delay times $\tau$, $T$, $t$, and the additional delay $\Delta$. The desired signal contains only contributions arising from the interactions with all five pulses. To ensure that the signal depends on the interaction of the ionization pulse, we assume that multi-photon ionization induced by the first four pulses is negligible, which can be achieved by appropriately adjusting their pulse intensities. Alternatively, a suitable chopping scheme could remove any $\Delta$-independent signal. Phase cycling of the relative phase differences between the first pulse phase $\phi_1$ and the phases of the second, third, and fourth pulses, $\phi_2, \phi_3, \phi_4$, ensures that only pathways containing at least one interaction with each of the first four pulses contribute to the signal. Detecting the nonlinear response arising from a single interaction with each phase-cycled pulse yields a signal equivalent to the four-wave mixing response measured in fourth-order F-2DES. Notably, this is usually not possible without ionization pulse as, dependent on the ionization potential of the system, additional interactions with the red four-pulse sequence would be necessary for ionization. Using the first four-pulse interactions, we prepare the same coherences and populations as in other optical 2DES techniques, involving the ground state $|g\rangle$, single-excited states $|\varepsilon_x\rangle$, and higher-excited states $|\alpha_x\rangle$. This is exemplarily illustrated in the energy-level scheme in Figure 1b for one possible excitation pathway starting from the ground state $|g\rangle$. While during the delay times $\tau$ and $t$, single-quantum coherences evolve (step 1 and step 3), single-exciton relaxation (step 2) and biexciton relaxation (step 4) occur during $T$ and $\Delta$, respectively.

Fourier transformations along $\tau$ and $t$ generate a correlation map between excitation and detection energies, indicating optical transitions between system states via diagonal peaks, while cross peaks indicate an interaction between two states, e.g., via coupling, energy transfer or EEA (Figure 1c). Additionally, the kinetic energy of the emitted photoelectrons can be resolved. Within this work, we consider a sufficiently weak excitation intensity of the four-pulse sequence such that higher-order contributions are negligible.[47] During the population time $T$, this limitation enables us to detect single-exciton dynamics without overlapping higher-excited-state dynamics. If higher excitation intensities are used such that higher-order terms start to contribute, they can be separated by adapting the concepts of phase cycling[4] and intensity cycling[48,49] for photoelectron detection. We leave a discussion of separation of higher orders in P-2DES to future work and concentrate here on the case that the response is dominated by the fourth-order nonlinearity arising from an individual interaction with each of the first four pulses, followed by ionization with the fifth (ultraviolet) pulse. If that ionization pulse projects the population of the system state occupied after the fourth pulse interaction directly into the detection channel, i.e., the photoemission yield, the system response is decoupled from dynamics occurring after the interaction with the fourth pulse, such as EEA. Consequently, the response exhibits characteristics like those in C-2DES, where the signal generation is not associated with population relaxation (Figure 1c, spectrum in the foreground). After four interactions with the four-pulse sequence, there is a possibility that higher excited states of the system will also be populated. Therefore, the ionization pulse delay time $\Delta$ introduces an additional scanning time parameter, which is sensitive to biexciton dynamics.

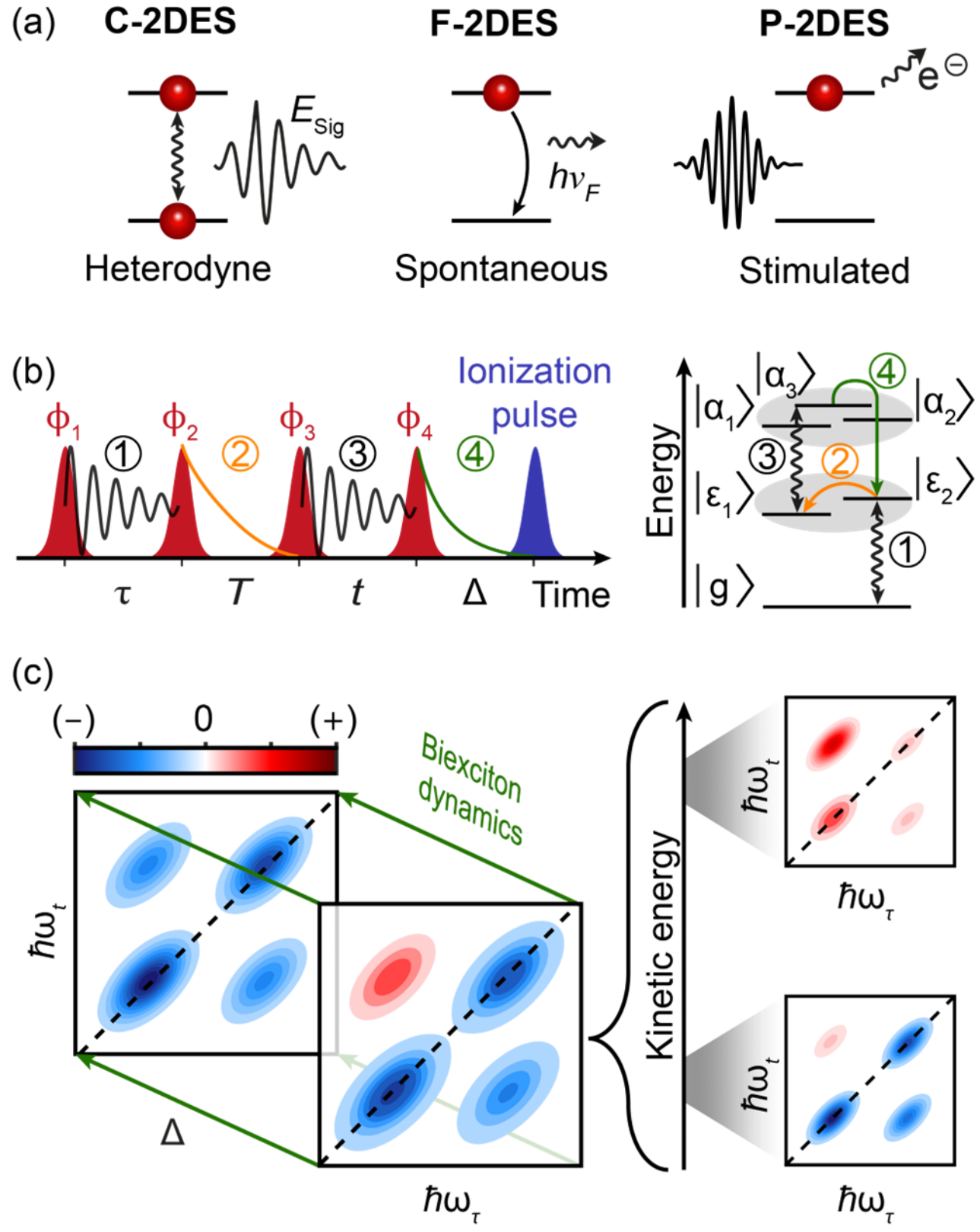

Figure 1. Ionization-pulse-assisted photoelectron-detected two-dimensional electronic spectroscopy. (a) Detection schemes of two-dimensional electronic spectroscopy (2DES). In coherently detected 2DES, the nonlinear electric field signal is typically retrieved via heterodyne detection with a local oscillator (left, C-2DES). In action-based techniques, population-based observables such as fluorescence photons are spontaneously emitted (center, F-2DES), while the emission of photoelectrons can be triggered by an ionization pulse (right, P-2DES). (b) Excitation scheme using a phase-cycled four-pulse sequence (red) followed by an additional ultraviolet ionization pulse (blue). An exemplary excitation pathway is illustrated in the energy-level scheme on the right. Wavy arrows indicate coherence between states, while solid arrows represent incoherent transfer of excitation energy. (c) Schematic absorptive 2D spectra of a molecular dimer obtained via ionization-pulse-assisted photoelectron emission. Left: A 2D spectrum at early delay time Δ is shown in the foreground, while a spectrum at later Δ is depicted in the background. The delay time between the fourth pulse and the ionization pulse (Δ) maps the biexciton dynamics. Right: The separation of the 2D spectrum into contributions at low and high kinetic energies, i.e., the photoemission from different states, illustrates how kinetic-energy filtering enables the disentanglement of the biexciton signature.

In the following, we discuss how single- and biexciton dynamics can be disentangled using P-2DES. We assume a photon energy of the ionization pulse that is insufficient to photoionize the ground state by one-photon absorption and a pulse intensity that is sufficiently low such that any contribution to the photoemission yield from multiphoton absorption is negligible. Using this assumption and an appropriate phase-cycling scheme, all signal contributions depend on four interactions with the four-pulse sequence and two subsequent interactions with the ionization pulse. Accordingly, the photoemission yield reflects the excited-state population at a time $\Delta$ following the fourth pulse. In the literature, the individual signal contributions are usually described in terms of Liouville pathways and visualized with double-sided Feynman diagrams.[46] The interaction with the ionization pulse introduces an additional factor $\Gamma_j$ for each diagram. The factor $\Gamma_j$ depends on the probability of photoionization of the state $|j\rangle$ that was populated prior to the interaction with the ionization pulse.

The delay time $\Delta$ is sensitive to an energy transfer between states if the transfer leads to a change of the detected total photoemission yield and therefore a change of the signal contribution of the corresponding Feynman diagram. For instance, in a two-level system, $\Delta$ probes the exciton lifetime, i.e., the relaxation from a single-excited state ($\Gamma_e := 1$) into the ground state ($\Gamma_g := 0$). In contrast, if two bound states $|n\rangle$ and $|m\rangle$ can be photoionized with identical probabilities, i.e., the corresponding Feynman diagrams exhibit the same $\Gamma_j$ value ($\Gamma_n = \Gamma_m$), a relaxation process during $\Delta$ from one of these two states into the other would not change the photoemission yield of a subsequent photoemission process. Therefore, this relaxation process would not be sampled by the total photoemission yield as a function of $\Delta$.

EEA is another example of a process that can change the final population prior to photoionization and the relative yield of the involved diagrams. This occurs when multiple excitons are simultaneously excited in a system, migrate, and annihilate upon encounter resulting in only one instead of two excitons. Commonly, EEA is described in a bimolecular model system that exhibits weak coupling and hence strongly localized states,[50] as illustrated in Figure 2. In this picture, we represent the weakly coupled system by two identical molecular sites, each with three electronic states: the ground state $|g\rangle$, the single-excited state $|e\rangle$, and the double-excited state $|f\rangle$. In the first step, the system is excited into a biexciton-state population, where both sites are in a single-excited state $|e\rangle$. In our case, the biexcitonic population is generated after four interactions of the four-pulse sequence with the system. In the second step, during the EEA process, the population transforms as the excitons can fuse into a configuration with one site in the double-excited state $|f\rangle$, while the other site relaxes to the ground state $|g\rangle$. This step can be described incoherently via Förster theory in the weak-coupling limit, or coherently as wave-packet motion when biexciton and double-excited states strongly mix.[51] The third step includes rapid internal conversion of the double-excited-state population, leading to a configuration in which one site is in the ground state $|g\rangle$ and the other site in a single-excited state $|e\rangle$.

Throughout this work, we utilize the picture of strongly localized states of EEA, i.e., we assume the weak-coupling limit. In this case, the dependence of the photoemission yield on $\Delta$ becomes apparent: At early delay times $\Delta$, one Liouville pathway ends in a configuration in which both sites are simultaneously excited and each of them can be ionized. Note that this does not describe a scenario in which two electrons are detected simultaneously, as the ionization pulse intensity is sufficiently weak to avoid multi-photon absorption. However, the probability of detecting one electron is enhanced due to the increased number of excited sites in comparison to a single-exited site and, consequently, $\Gamma_{ee} > \Gamma_e$ (Figure 2, dashed and solid blue arrows, left). If $\Delta$ is increased, energy transfer from one site to the other can occur before the photoemission process takes place. After the energy transfer, one site is in the ground state and the other site in a double-excited state. Given that the ionization-pulse photon energy and the pulse intensity are insufficient to depopulate the ground state, only the ionization of the double-excited site contributes to the signal yield (Figure 2, one blue arrow, center). If $\Delta$ is increased further, relaxation may occur from the double-excited state to the single-excited state, followed by ionization (Figure 2, one blue arrow, right). In contrast to the first case (early delay times $\Delta$, Figure 2 left), only one site resides in the excited state during the second and the third case (Figure 2, middle and right). Statistically, fewer excited sites lead to a decrease in photoemission yield relative to the initially excited biexciton state. For two independent monomers, the signal contribution of a diagram ending in the population with both sites in the single-excited state prior to photoionization (factor $\Gamma_{ee}$) would be two times larger than for the scenario that only one site is in the single-excited state (factor $\Gamma_e$). Note that the ratio $\Gamma_{ee}/\Gamma_e$ can differ from 2 in a coupled dimer system since the exact value depends on the corresponding photoemission transition matrix elements. However, as long as $\Gamma_{ee} \neq \Gamma_e$, EEA leads to a $\Delta$-dependent evolution of the total photoelectron yield.

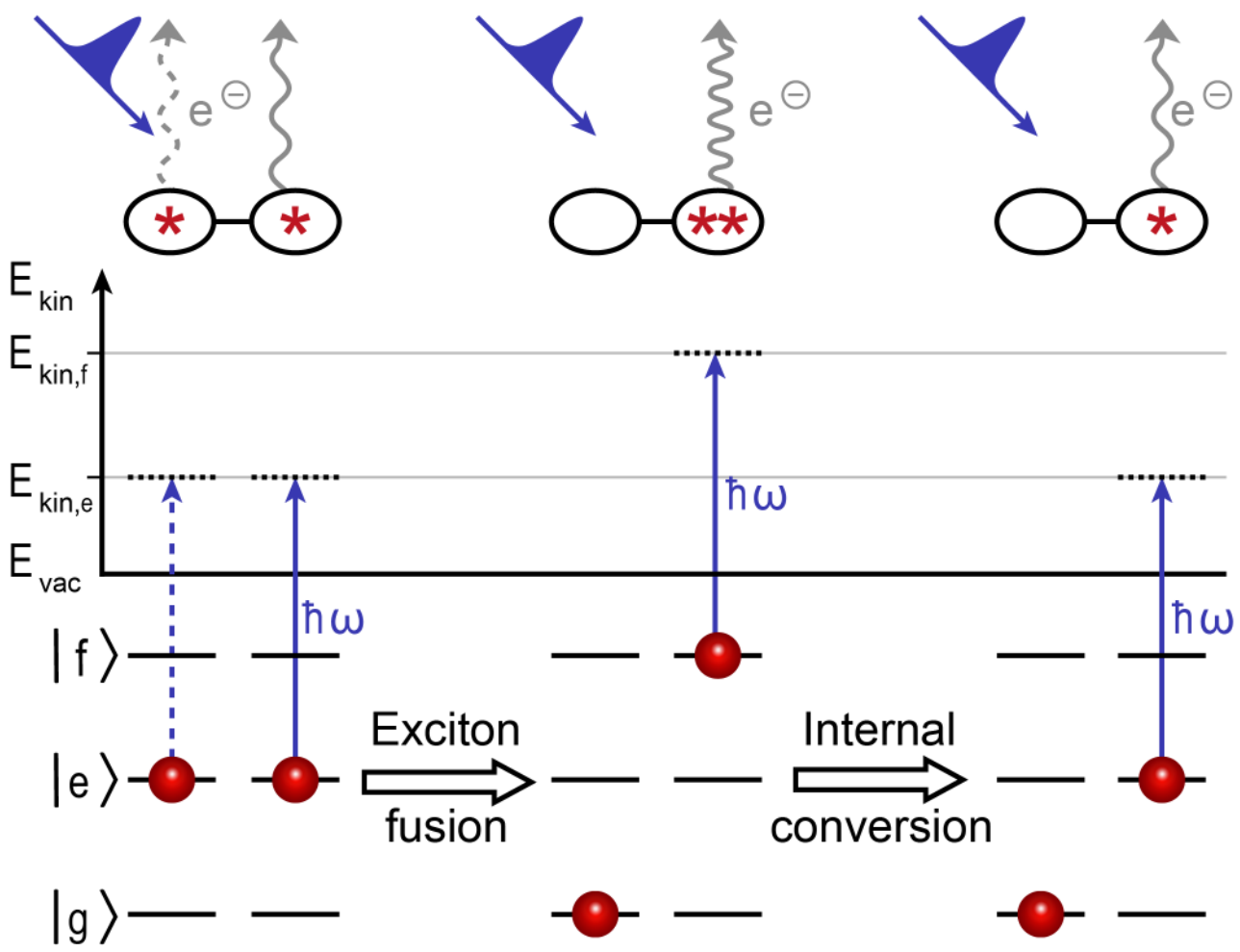

Figure 2. Site-basis picture of exciton–exciton annihilation steps and corresponding photoemission yield. The photoemission process is shown for three possible stages during annihilation dynamics in a homodimer. Initially, both monomers are in their respective single-excited states. Both can interact independently with the ionization pulse, resulting in the emission of a single electron from either one monomer (solid arrow) or the other (dashed arrow) with kinetic energy contributing to $E_{\mathrm{kin},e}$. For simplicity only a single ionization channel for each excited state manifold is illustrated. If the photoemission process does not occur, exciton fusion into one double-excited state $|f\rangle$ can take place in a second step. At this stage, the interaction with the ionization pulse leads to photoemission of an electron with kinetic energy contributing to $E_{\mathrm{kin},f}$. If this photoemission process does not occur either, internal conversion leads to one single-excited monomer in a third step. Photoemission at this final stage generates a photoelectron with kinetic energy contributing to $E_{\mathrm{kin},e}$.

Beyond detecting the total photoelectron yield, additional information can be obtained by resolving the photoemission signal with respect to electron kinetic energy. In the localized two-site basis set, each molecule can be photoionized, which may occur from either the single- or the double-excited configuration. Within the frozen orbital approximation, Koopmans' theorem yields an estimate for the ionization potentials of these states, which in turn determine the photoelectron kinetic energy features associated with photoionization from $|e\rangle$ and $|f\rangle$ manifolds, respectively. In general, photoionization from a population in the single-excited manifold can give rise to multiple features in the kinetic energy spectrum, depending on accessible cationic final states. Accordingly, we define the decomposed kinetic energy spectra $E_{\mathrm{kin},e}$ and $E_{\mathrm{kin},f}$, which contain only contributions arising exclusively from photoionization out of single-excited and double-excited site state populations, respectively. Together, $E_{\mathrm{kin},e}$ and $E_{\mathrm{kin},f}$ provide the full kinetic energy spectrum. The features of $E_{\mathrm{kin},e}$ and $E_{\mathrm{kin},f}$ may partially or fully overlap.

During the EEA process (Figure 2), the double-excited state is populated in the second step and experiences population decay in the third step. If at least one feature of $E_{\mathrm{kin},f}$ can be spectrally separated from the features in $E_{\mathrm{kin},e}$, the EEA process leads to a change of the kinetic energy of the detected electrons during the EEA process. This opens a way to directly follow the annihilation pathway in time by kinetic-energy filtering as we will discuss below. Whether ionization from higher- and lower-lying excited states leads to spectrally separable contributions in the kinetic energy distribution is system dependent. The energy-filtering approach presented below is applicable in cases where at least one feature in the kinetic energy spectrum can be predominantly attributed to photoemission from double-excited (i.e., $f$-like) states. Such conditions are, for example, found in conjugated systems such as polyenes,[52] and some benzene derivatives,[53,54] where distinct photoelectron features associated with different excited states have been reported.

For modeling the photoinduced molecular dynamics as well as the photoemission process, we numerically solve the Lindblad master equation:

$$\frac{\partial\hat{\rho}(t)}{\partial t} = -\frac{i}{\hbar}\left[\hat{H}(t),\hat{\rho}(t)\right] + L_{\mathrm{decay}} + L_{\mathrm{PE}}. \qquad (1)$$

Here, $\hat{\rho}(t)$ and $\hat{H}(t)$ denote the system's density matrix and Hamilton operator, respectively. The Lindbladian $L_{\mathrm{decay}}$ describes $k$ decay channels which represent population relaxation or pure dephasing between bound system states $|i\rangle$ and $|j\rangle$:

$$L_{\mathrm{decay}} = \sum_k \frac{1}{\tau_k}\left(\hat{\mathcal{L}}_k\hat{\rho}(t)\hat{\mathcal{L}}_k^\dagger - \frac{1}{2}\hat{\mathcal{L}}_k^\dagger\hat{\mathcal{L}}_k\hat{\rho}(t) - \frac{1}{2}\hat{\rho}(t)\hat{\mathcal{L}}_k^\dagger\hat{\mathcal{L}}_k\right). \tag{2}$$

Here, the jump operator is $\hat{\mathcal{L}}_k = \hat{\sigma}_{ij} = |i\rangle\langle j|$ in case of population relaxation and $\hat{\mathcal{L}}_k = |i\rangle\langle i| + |j\rangle\langle j|$ for a pure dephasing process, while $\tau_k$ denotes the corresponding phenomenological time constant. Finally, we model photoelectron emission as an incoherent energy transfer process via the term $L_{\mathrm{PE}}$ on the right side of equation (1):

$$L_{\mathrm{PE}} = \sum_{k'} \gamma_{\mathrm{PE}_{k'}} |E^+(t)|^2 \left(\hat{\mathcal{L}}_{k'}\hat{\rho}(t)\hat{\mathcal{L}}_{k'}^\dagger - \frac{1}{2}\hat{\mathcal{L}}_{k'}^\dagger\hat{\mathcal{L}}_{k'}\hat{\rho}(t) - \frac{1}{2}\hat{\rho}(t)\hat{\mathcal{L}}_{k'}^\dagger\hat{\mathcal{L}}_{k'}\right). \tag{3}$$

Using jump operators $\hat{\mathcal{L}}_{k'} = \hat{\sigma}_{P_i j} = |P_i\rangle\langle j|$ for the $k'$-th photoemission channel, equation (3) takes the form of a Lindbladian in which population is transferred from a bound state $|j\rangle$ to a free-electron state $|P_i\rangle$ that lies energetically above the ionization potential. The overall rate constant of this transfer process is determined by two quantities: The first one is the time-independent constant $\gamma_{\mathrm{PE}_{k'}}$, which is a measure for the photoemission matrix element of the $|j\rangle$-to-$|P_i\rangle$ transition. The second quantity, i.e., $|E^+(t)|^2$, is time dependent and scales with the slowly varying intensity of the complex-valued light field $E^+(t)$ interacting with the system, where the real-valued electric field is $E(t) = E^+(t) + E^-(t)$ and $E^-(t)$ is the complex conjugate of $E^+(t)$. Here, when defining the $k'$-th photoemission channel, $|E^+(t)|^2$ represents the intensity of the ionization pulse since we assume a weak excitation limit for the four-pulse sequence.

The Liouville equation with the Lindbladian $L_{\mathrm{decay}}$ can be numerically propagated with our already existing QDT software package,[45] while the population transfer describing photoemission via $L_{\mathrm{PE}}$ was added to the package during this work. In doing so, the total electron emission yield $Y$ is calculated for each pulse sequence by summing over all populations of free-electron states $|P_i\rangle$ at a single point of time $t'$, where a stationary regime is reached (illustrated in Figure S1e and f). The free-electron states $|P_i\rangle$ hence serve as containers for the photoemission yield and can be assigned a specific kinetic energy features when kinetic-energy resolution is required. In our simulations each excited eigenstate $|j\rangle$ of the system is connected to a single representative free-electron state $|P_j\rangle$, consistent with the above-mentioned assumption that multiphoton ionization processes are neglected. Further details about the implementation of photoemission into QDT are given in Section S1 of the Supporting Information.

As a model system, we consider a weakly coupled J-type heterodimer (see Supporting Information Table S1 for system parameters), consisting of two three-level subsystems, each with a ground state $|g\rangle$, a single-excited state $|e\rangle$, and a double-excited state $|f\rangle$. The eigenstates of the system in the exciton basis were obtained by expanding the Hilbert subspaces of the molecular entities to the dimension of the total Hilbert space and diagonalizing the overall Hamiltonian. Eigenstates describing the excited states are labeled by Greek letters according to the energy level scheme in Figure 1b. The weak coupling between the two sites results in minimal excitonic delocalization, allowing the annihilation and photoemission processes to be described in the localized framework introduced above. This approach connects the collective states excited by the four-pulse sequence and the single-particle states involved in the photoemission process. In our model, the eigenstates $|\varepsilon_1\rangle$ and $|\varepsilon_2\rangle$ are strongly localized at a single site and are composed of the site basis states $|e_A\rangle$ and $|e_B\rangle$. Similarly, if the system is in one of the eigenstates $|\alpha_1\rangle$ or $|\alpha_2\rangle$, predominantly one site is in the ground state while the other site is in the double-excited state $|f_A\rangle$ or $|f_B\rangle$. In contrast, eigenstate $|\alpha_3\rangle$ has strong biexciton character, where both sites are in their single-excited state ($|e_A e_B\rangle$). The absolute square of the transformation coefficients, which give the weights of the site basis states in each higher-excited eigenstate of the exciton basis, are listed in the Supporting Information Section S2. The implications of the weak coupling limit are discussed in Section S3 of the Supporting Information.

The rate constant $\gamma_{\mathrm{PE}_{k'}}$ of the photoemission Lindbladian and, consequently, also the factor $\Gamma_j$ of a specific Liouville pathway, depend on the sample and the specific experimental parameters. Here, we start with the simplest case, where all factors containing states with strong localized character are the same, i.e., $\Gamma_{\alpha_2} = \Gamma_{\alpha_1} = \Gamma_{\varepsilon_2} = \Gamma_{\varepsilon_1} := 1$. If these factors are not equal, additional ESA-type contributions can appear in the 2D spectra (Supporting Information Section S4). We discuss below and in the Supporting Information (Section S5, and Section S6) scenarios with varied factors, their influence on the $\Delta$-dependent dynamics and how signatures of EEA could still be extracted. As described above, photoemission from an excited state with strong biexciton character is expected to exhibit a larger factor $\Gamma_j$ than from a localized state, since photoionization can occur from two excited sites rather than only one. For the population $|e_A e_B\rangle\langle e_A e_B|$, ionization may proceed via site "A", or site "B", or both sites simultaneously, corresponding to the matrix elements $|P_A e_B\rangle\langle P_A e_B|$, $|e_A P_B\rangle\langle e_A P_B|$, and $|P_A P_B\rangle\langle P_A P_B|$, respectively. The latter process is neglected here due to the weak ionization pulse amplitude. Tracing over the remaining ionization channels yields the effective rate $\Gamma_{e_A e_B}$. In the weak coupling limit, $\Gamma_{\alpha_3} \approx \Gamma_{e_A e_B}$, $\Gamma_{\varepsilon_1} \approx \Gamma_{e_B}$, $\Gamma_{\varepsilon_2} \approx \Gamma_{e_A}$ leading to the relation $\Gamma_{\alpha_3} = \Gamma_{\varepsilon_2} + \Gamma_{\varepsilon_1}$. In the following we set $\Gamma_{\alpha_3} = \Gamma_{\varepsilon_2} + \Gamma_{\varepsilon_1} = 2$, while alternative scenarios are discussed in Supporting Information Section S6. Although the exact values depend on the photoemission matrix elements, the main observations presented below hold for other values as long as $\Gamma_{\alpha_3} \neq \Gamma_{\varepsilon_1}$. In our model system, photoionization from $|\alpha_1\rangle$ and $|\alpha_2\rangle$, primarily characterized by a doubly excited site, yields emitted electrons with higher kinetic energy than photoionization from states $|\varepsilon_1\rangle$, $|\varepsilon_2\rangle$, and $|\alpha_3\rangle$, where predominately neither of the two sites is in the double-excited state.

We model the second step of annihilation as an incoherent energy transfer from $|\alpha_3\rangle$ to $|\alpha_1\rangle$ or to $|\alpha_2\rangle$, i.e., biexciton relaxation, with an effective decay time constant of 150 fs, followed by a faster relaxation with a time constant of 50 fs to $|\varepsilon_1\rangle$ or $|\varepsilon_2\rangle$. Additionally, an incoherent energy transfer from $|\varepsilon_2\rangle$ to $|\varepsilon_1\rangle$, i.e., single-exciton relaxation, is considered with a time constant of 250 fs. In this work, we focus on dynamics occurring on short time scales of up to 1 ps delay, which are much shorter than the typical exciton lifetime in the nanosecond regime. To ensure computational efficiency and reduce simulation time, the exciton lifetime is therefore not explicitly included in our model and appears infinitely long in comparison to all other time constants.

Figure 3a shows four representative P-2DES spectra corresponding to the different combinations of two selected early (20 fs) and late (500 fs) $T$ and $\Delta$ delays. Here, ground-state bleach (GSB) and stimulated emission (SE) signals are negative, and excited-state absorption (ESA) contributions are positive. All 2D maps exhibit four peaks arranged in a square pattern, with two negative diagonal peaks and their corresponding cross peaks. Additional ESA signals arising from coherences between the eigenstates of the single-excited-state manifold $\{|\varepsilon_1\rangle, |\varepsilon_2\rangle\}$ and those of the double-excited manifold $\{|\alpha_1\rangle, |\alpha_2\rangle\}$ with strong localized character are absent, as previously shown for action-detected 2DES.[19] Note that additional peaks appear if $\Gamma_{\alpha_x} \neq \Gamma_{\varepsilon_x}$ with $x \in \{1,2\}$ (see Supporting Information Section S4). At early delay times $T$ and $\Delta$ (Figure 3a, A), the 2D map shows the characteristic peaks of a J-type coupled dimer as it would be measured via C-2DES.[20] Figure S8a and b show an explicit comparison of a C-2DES spectrum at $T = 20$ fs and the spectrum shown in Figure 3a, for $T = 20$ fs and $\Delta = 20$ fs, respectively. Here, oscillator strength redistribution enhances the amplitude of the low-energy diagonal peak relative to the high-energy diagonal peak. The cross peak above the diagonal is positive, while the cross peak below the diagonal is negative, revealing the character of the J-type coupling. As in C-2DES, the single-exciton relaxation from $|\varepsilon_2\rangle$ to $|\varepsilon_1\rangle$ can be observed as a decrease in the absolute magnitude of the high-energy diagonal peak and a simultaneous increase of the absolute magnitude of the cross-peak amplitude below the diagonal while scanning $T$ (Figure 3b). The solid lines are exponential fits with a time constant of 250 fs for single-exciton relaxation. For the fit, we have excluded data points for population times <150 fs where interexcitonic coherences dominate and distort the signal amplitude in addition to pulse overlap effects. Notably, the signature of the interexcitonic coherence evolving during $T$ is stronger at late $\Delta$ than at early $\Delta$ (Figure S9). Similarly, an enhanced sensitivity to such coherent excited-state dynamics was proposed for F-2DES compared to C-2DES.[55] At short $T$ and late $\Delta$ delays (Figure 3a, B), the 2D map exhibits features typical for F-2DES spectra (Figure S8c). In contrast to 2D maps for early $\Delta$, the coupling signatures, i.e., the cross peaks with alternating signs, are not present; instead, both cross peaks are negative, indicating an EEA process during $\Delta$. As described above, the EEA process can also lead to a substantial static background, reducing the contrast of the energy-transfer signal as a function of $T$. As a result, an accurate determination of the energy transfer rate could become challenging in F-2DES, especially for large systems.[6,24]

In P-2DES, it is possible to control the amount of static background arising from EEA after the interaction with the four-pulse sequence by adjusting the delay $\Delta$. For small delays $\Delta$, the system does not have sufficient time to undergo EEA before the projection of state populations into the detection channel by the ionization pulse. For larger delays, EEA occurs prior to photoemission resulting in a weaker spectral signature of energy transfer. We observe this effect by the change in relative peak amplitude of the diagonal peak and the cross peak at $\hbar\omega_{\tau} = 1.89$ eV between $T = 20$ fs and $T = 500$ fs which is stronger for $\Delta = 20$ fs (Figure 3a, A and C) than for $\Delta = 500$ fs (Figure 3a, B and D). Therefore, as proposed by Bolzonello et al.,[24] the extent of incoherent mixing leading to static background can be reduced by time gating in action-detected 2DES of a dimer. Furthermore, scanning $\Delta$ enables one to investigate EEA. Figure 3c shows that all peak amplitudes decrease as a function of $\Delta$ with the biexciton relaxation time of 150 fs. The strongest effect is present for the cross peak above the diagonal because of the oscillator strength redistribution resulting from J-type coupling. At late $\Delta$ delays, the absolute magnitude of all peak amplitudes would decay to zero due to finite lifetimes, as shown by Malý and Mančal.[22] Here, we concentrate on the early time scales and thus set the exciton lifetime to infinity. As we set $\gamma_{\varepsilon_2} = \gamma_{\varepsilon_1}$, the single-exciton energy transfer taking place during $\Delta$ does not lead to a change of the transient photoemission signal. Therefore, the single-exciton and biexciton dynamics can be investigated individually by scanning the delay times $T$ and $\Delta$, respectively. In the general case where $\gamma_{\varepsilon_2} \neq \gamma_{\varepsilon_1}$, both EEA and single-exciton energy transfer alter the photoelectron signal as a function of $\Delta$. Nevertheless, it is possible to disentangle both contributions by performing a scan of $T$, which provides the single-exciton dynamics. The resulting information enables the extraction of the EEA dynamics from the $\Delta$-scan, as demonstrated in Section S6.

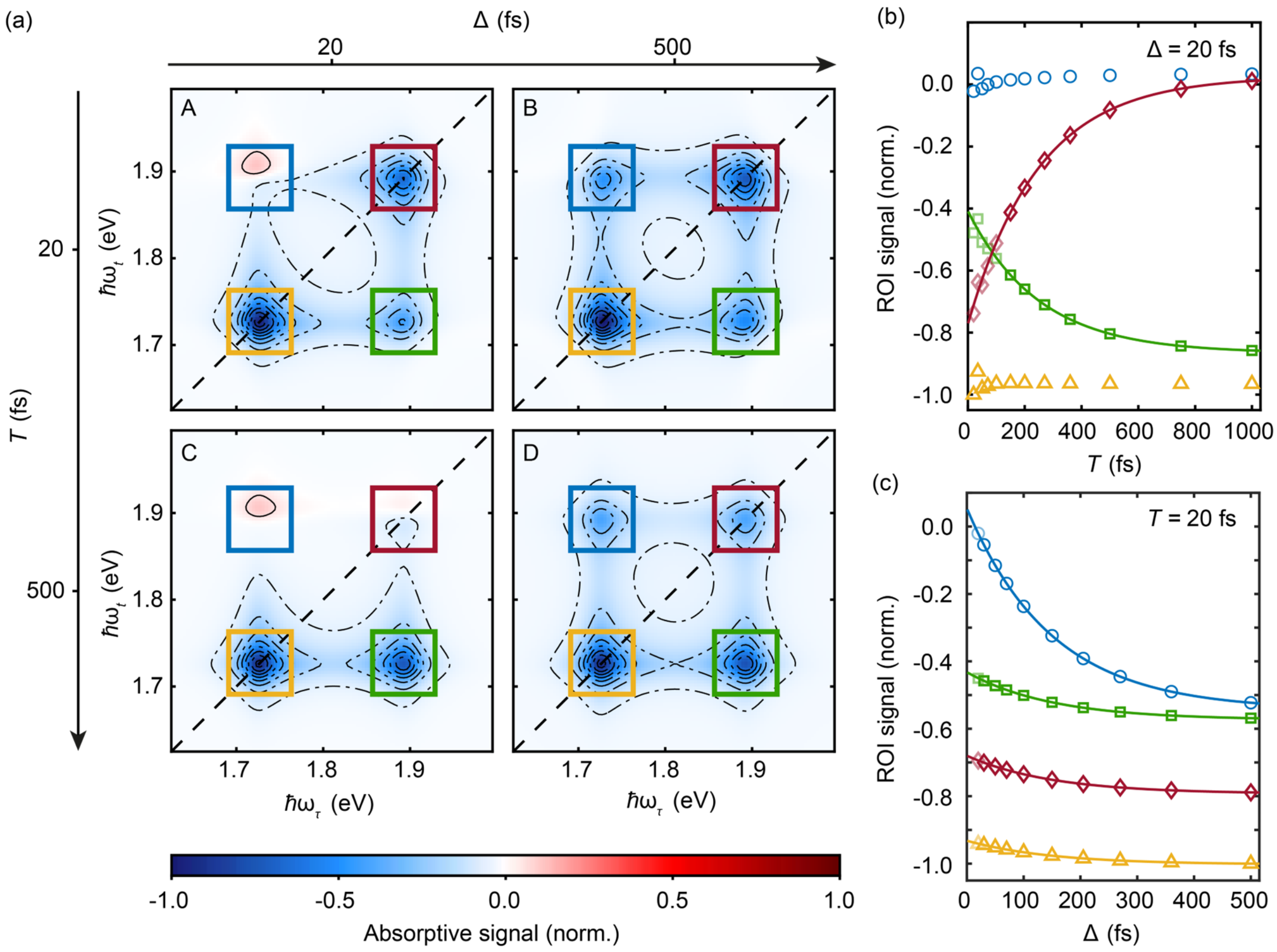

Figure 3. Simulated photoelectron-detected 2D spectra of a heterodimer. (a) Absorptive 2D spectra for four combinations (A–D) of time delays $T$ and $\Delta$. All four 2D spectra are normalized to the global maximum absolute value. The dashed diagonal line is drawn along $\hbar\omega_\tau = \hbar\omega_t$. Each square marks a region of interest (ROI) chosen to monitor the dynamics during $T$ or $\Delta$. (b) Evolution of integrated signals of ROIs as a function of time delay $T$ at constant $\Delta = 20$ fs (symbols). (c) Evolution of ROI signals as a function of time delay $\Delta$ at constant $T = 20$ fs (symbols). Solid lines in panels b and c depict exponential fits. Semitransparent symbols represent values excluded from the corresponding fit due to pulse-overlap effects and electronic coherences.

Recent studies focused on the cross peaks in F-2DES and the influence of EEA that obscures signatures of coupling[18,19,23] and energy transfer.[6,24] In Figure 3c, however, we observe biexciton dynamics in form of EEA that is seen as an increase of the absolute amplitudes of all peaks with the biexciton relaxation time of 150 fs. Especially for the high-energy diagonal peak (Figure 3a, red square, $\hbar\omega_\tau = 1.89$ eV, $\hbar\omega_t = 1.89$ eV), we observe a significantly stronger negative feature for late $\Delta$ delays ($\Delta = 500$ fs) than for early $\Delta$ delays ($\Delta = 20$ fs) at late $T$ delays. Additionally, focusing on the short time delays $T$ (Supporting Information, Section S8) reveals a suppressed signature of interexcitonic coherences at early $\Delta$ delays. In contrast, after efficient EEA (at late $\Delta$ delays) the spectral signatures of interexcitonic coherences become stronger and lead to a significant oscillating signal at the diagonal peak positions (Figure S9).

In the following, we analyze the origin of the enhanced interexcitonic coherences and pronounced EEA feature of the high-energy diagonal peak. As described above, QDT numerically solves the Lindblad master equation and does not rely on a perturbative expansion including double-sided Feynman diagrams. Nevertheless we utilize double-sided Feynman diagrams (Figure 4), which are a versatile tool for dissecting the photoexcitation processes underlying multi-dimensional spectroscopy,[56,57] to discuss the properties of the high-energy diagonal peak. For simplicity, we assume strict time ordering and impulsive interactions in all double-sided Feynman diagrams, although the signals simulated with QDT are not restricted to these assumptions. All diagrams can be classified as GSB, SE, or ESA. If an ESA diagram exhibits a biexciton population after the fourth interaction, it is labeled with ESA2 and exhibits a positive sign in our convention, which is opposite to the minus sign of all other contributions. In the limit of a single interaction per phase-cycled pulse, these ESA2 pathways are responsible for all EEA signatures since higher-excited states are populated in ESA2 pathways only. At cross-peak positions, ESA2 pathways always contribute, thereby obscuring the coupling features as discussed earlier.[20,22,55] In contrast, an ESA2 pathway only contributes to the diagonal peaks if either an interexcitonic coherence or a single-exciton transfer is present during $T$. We demonstrate this exemplarily in Figure 4 for the high-energy diagonal peak ($\hbar\omega_\tau = 1.89$ eV, $\hbar\omega_t = 1.89$ eV), corresponding to the energy gaps between $|g\rangle$ and $|\varepsilon_2\rangle$ as well as between $|\varepsilon_1\rangle$ and $|\alpha_3\rangle$. Other eigenstates are not directly excited at this spectral position but contribute as intermediates of the biexciton relaxation process. All shown diagrams have nonrephasing phase signature $\phi = +\phi_1 - \phi_2 + \phi_3 - \phi_4$. Corresponding rephasing diagrams which can be obtained by interchanging the first two interactions are listed in the Supporting Information Section S10. Time evolves from bottom to top. If the final population before ionization relaxes into another state, the pathway is indicated with a prime. For simplicity, "late" time delays refer to those at which all populations have relaxed to the lowest-lying excited state. All ESA2 pathways depend on the same transition dipole moments as the ESA pathways to their left, and thus they interfere destructively with each other.

The choice of the delays $\Delta$ and $T$ has an influence on the contrast of the observed processes evolving during $T$ and $\Delta$, respectively. We first discuss the presence of interexcitonic coherences. At early $T$ and $\Delta$ delays, the nonrephasing signal is given by the contributions of GSB, SE, SE$_C$, ESA$_C$, and ESA2$_C$ (Figure 4, top left). Due to EEA, the ESA2$_C$ pathway contribution vanishes at late $\Delta$ delays and is replaced by ESA2$_C$' (Figure 4, top right). The ESA2$_C$'and ESA$_C$' pathways cancel each other as they share not only the same transition dipole moments but also the same $\Gamma_j$ due to the same population ($|\varepsilon_1\rangle\langle\varepsilon_1|$) before photoionization occurs. Because of the cancellation of both ESA-type pathways at late $\Delta$, the SE$_C$ pathway contributes exclusively to the total signal leading to the oscillating signature of the interexcitonic coherence. This is not the case for early $\Delta$ delays if the ionization probability from the state with strong biexcitonic character ($|\alpha_3\rangle$) is higher than the one of the single-exciton state $|\varepsilon_1\rangle$. In that case, the pathways ESA$_C$ and ESA2$_C$ only partially cancel each other, leaving a remaining positive signal contribution which destructively interferes with SE$_C$. Therefore, the signature of interexcitonic coherences is weakened at early $\Delta$. In case of $\mu_{g\varepsilon_2}\mu_{g\varepsilon_1} - \mu_{\varepsilon_2\alpha_3}\mu_{\varepsilon_1\alpha_3} = 0$ and a ratio of $\Gamma_{\alpha_3}/\Gamma_{\varepsilon_1} = 2$, all coherence pathways mutually cancel, suppressing the signature of interexcitonic coherences during $T$ at early $\Delta$ entirely. Even in more general cases, the interexcitonic coherence contributions are initially weak (early $\Delta$) and become significant at larger $\Delta$ delays, i.e., after annihilation took place efficiently. As a consequence, an efficient annihilation process during $\Delta$ leads to an enhanced contrast of the interexcitonic coherence dynamics as a function of $T$.

Let us now focus on the signatures of biexciton dynamics that can be resolved as a function of $\Delta$. In the limit of weak excitation intensities, only pathways ending in a higher-excited-state population before interaction with the ionization pulse contain signatures of biexciton dynamics. At early $T$ delays, the signals for early and late $\Delta$ are described by the upper left and upper right sets of diagrams in Figure 4, respectively. The replacement of the ESA2C pathway with ESA2C' is responsible for the EEA dynamics of the diagonal peak (Figure 3c, red line). Due to the (partial) cancellation of the coherence pathways (i.e., pathways with subscript "C") described above and their fast dephasing, the EEA signature as a function of $\Delta$ is weak and lies on top of the static GSB and SE signal contributions. The situation is different for late $T$, causing a strong contrast of the EEA dynamics as a function of $\Delta$ (Figure S10). Here, the coherence pathways present at early $T$ (Figure 4, upper left and upper right) dephase and no longer contribute to the overall signal, including all ESA-type pathways. The strong EEA signature as a function of $\Delta$ of the high-energy diagonal peak at late $T$ is caused by single-exciton energy transfer during $T$: The SE pathways present at early $T$ (Figure 4, top) are replaced by two oppositely signed ESA contributions (ESA and ESA2) (Figure 4, bottom). At early $\Delta$, the energy transfer leads to a strongly decreased absolute magnitude of the signal because the positive ESA2 and the two negative contributions GSB and ESA partially interfere destructively. As a consequence, the static background is reduced at late $T$. For late $T$ and $\Delta$ delays, the diagonal peak re-emerges because EEA replaces the ESA2 (early $\Delta$) with the ESA2' (late $\Delta$), which leads to the cancellation of the two ESA-type pathways, and GSB' remains. In summary, the contrast of the biexciton dynamics as a function of Δ is stronger for the high-energy diagonal peak at late $T$ than at early $T$ due to single-exciton energy transfer.

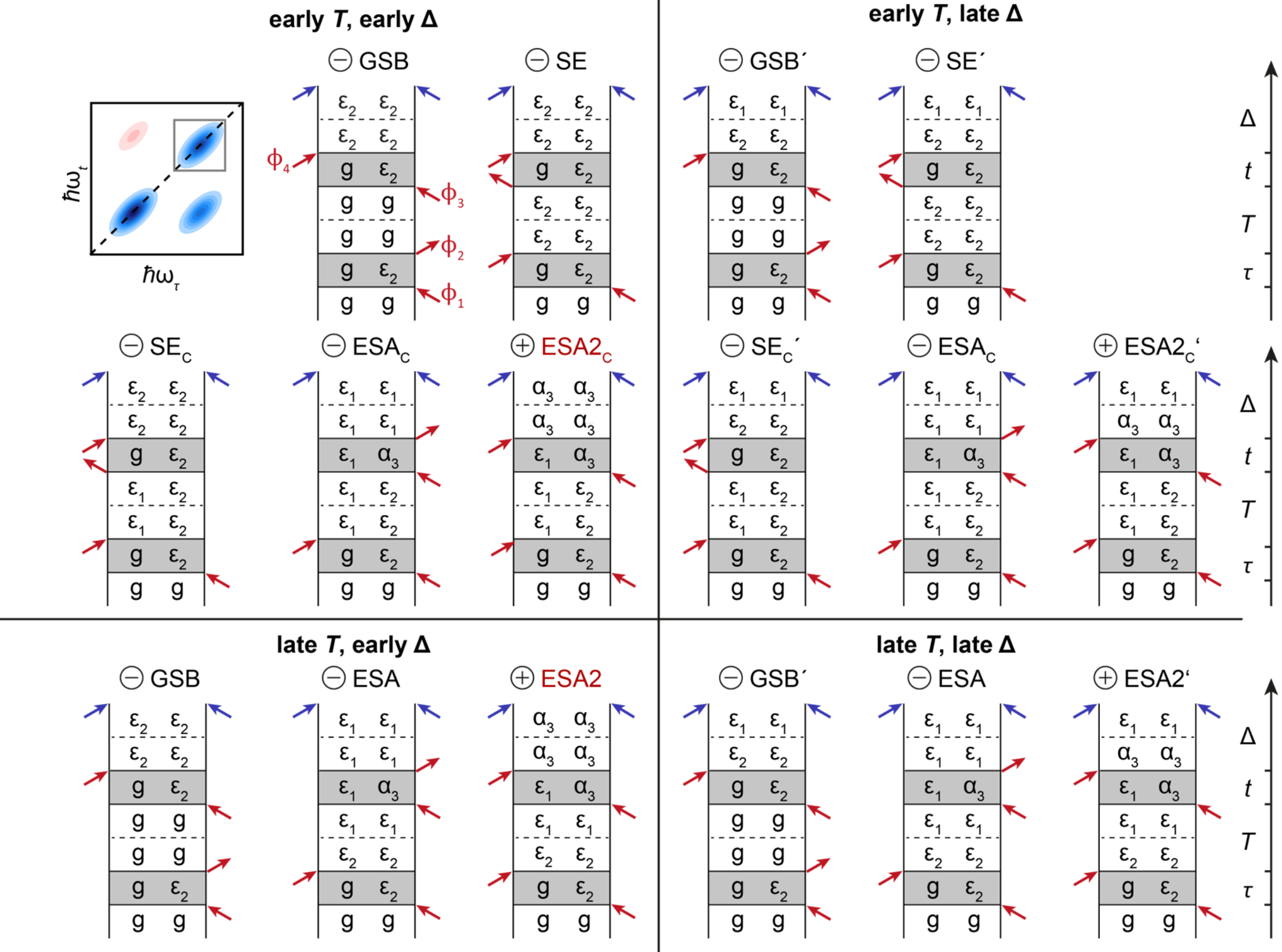

Figure 4. Nonrephasing Liouville pathways of the high-energy diagonal peak as depicted in the scheme on the top left corner. Four different cases are distinguished for the selection of delay times $T$ and Δ. Their groupings, separated by the vertical and horizontal black lines, correspond to the arrangement of data panels in Figure 3a according to early and late delay times $T$ and Δ. Time ordering of all pulses is assumed with respect to the sequence shown in Figure 1b. Red arrows represent interactions with the phase-cycled four-pulse sequence while blue arrows indicate interactions with the ionization pulse. Phase labels are only illustrated for the first diagram and are removed for all others for readability. Characters on the left-hand side of each diagram represent the ket state vectors; characters on the right-hand side represent the bra state vectors. The sign and type of each diagram are indicated above. If pathways exhibit a coherence during $T$, they are marked by the subscript "C"; if a pathway includes relaxation after the fourth pulse it is labeled with a prime. All pathways contribute with $\Gamma_j = 1$, except the red-marked ESA2 pathways, which contribute with $\Gamma_{\alpha_3} = 2$.

A feature of P-2DES is its capability to resolve the kinetic energy of photoelectrons. While the time-gating approach captures only energy transfer between states exhibiting a different ionization probability upon photoionization, resolving the kinetic energy of photoelectrons provides a direct temporal tracking of different state populations. As described above, in the weak-coupling limit the kinetic energy features contained in $E_{\mathrm{kin},e}$, associated with the photoemission of single-excited site states, can be assigned to photoemission from exciton-basis states $|\varepsilon_1\rangle$, $|\varepsilon_2\rangle$, and $|\alpha_3\rangle$. Photoemission from states $|\alpha_1\rangle$ and $|\alpha_2\rangle$ lead to kinetic energy features in $E_{\mathrm{kin},f}$. In our example, photoemission from $|\alpha_1\rangle$ and $|\alpha_2\rangle$ leads to features at higher kinetic energies than photoemission from $|\varepsilon_1\rangle$, $|\varepsilon_2\rangle$, and $|\alpha_3\rangle$. Figure 5a shows a kinetic-energy-filtered 2D spectrum, which is calculated by the fraction $Y'$ of the total electron emission yield $Y$. The fraction $Y'$ contains the sum over all final populations of free-electron states $|P_j\rangle$ that are connected to photoemission from states $|\alpha_1\rangle$ and $|\alpha_2\rangle$. Only positive peaks are present, as solely those pathways contribute here that end in a higher-excited state after the system's interaction with the fourth pulse. This condition is only fulfilled by ESA2-type pathways, all of which exhibit a positive sign. The spectrum is dominated by a cross peak above the diagonal. This feature arises due to an excitation of state $|\alpha_3\rangle$ with the fourth pulse interaction followed by a relaxation into $|\alpha_2\rangle$ and $|\alpha_1\rangle$. This relaxation process leads to an increasing high-kinetic-energy signal at early $\Delta$ (Figure 5b, blue). The subsequent decay of the signal beyond 100 fs reflects further relaxation into the single-excited states from states $|\alpha_2\rangle$ and $|\alpha_1\rangle$. In contrast, at $\hbar\omega_t = 1.71$ eV, corresponding to the energy gap between $|\varepsilon_1\rangle$ and $|\alpha_2\rangle$, the signal decays rapidly without delay (Figure 5b, purple). Here, the direct population of $|\alpha_2\rangle$ dominates the signal, followed by rapid relaxation to the single-exciton manifold without an additional transfer step as in the other peak. Figure S4 shows a 2D spectrum simulated without relaxation from $|\alpha_3\rangle$ and emphasizes spectral regions dominated by direct decay versus those where annihilation from the biexciton channel dominates the dynamics.

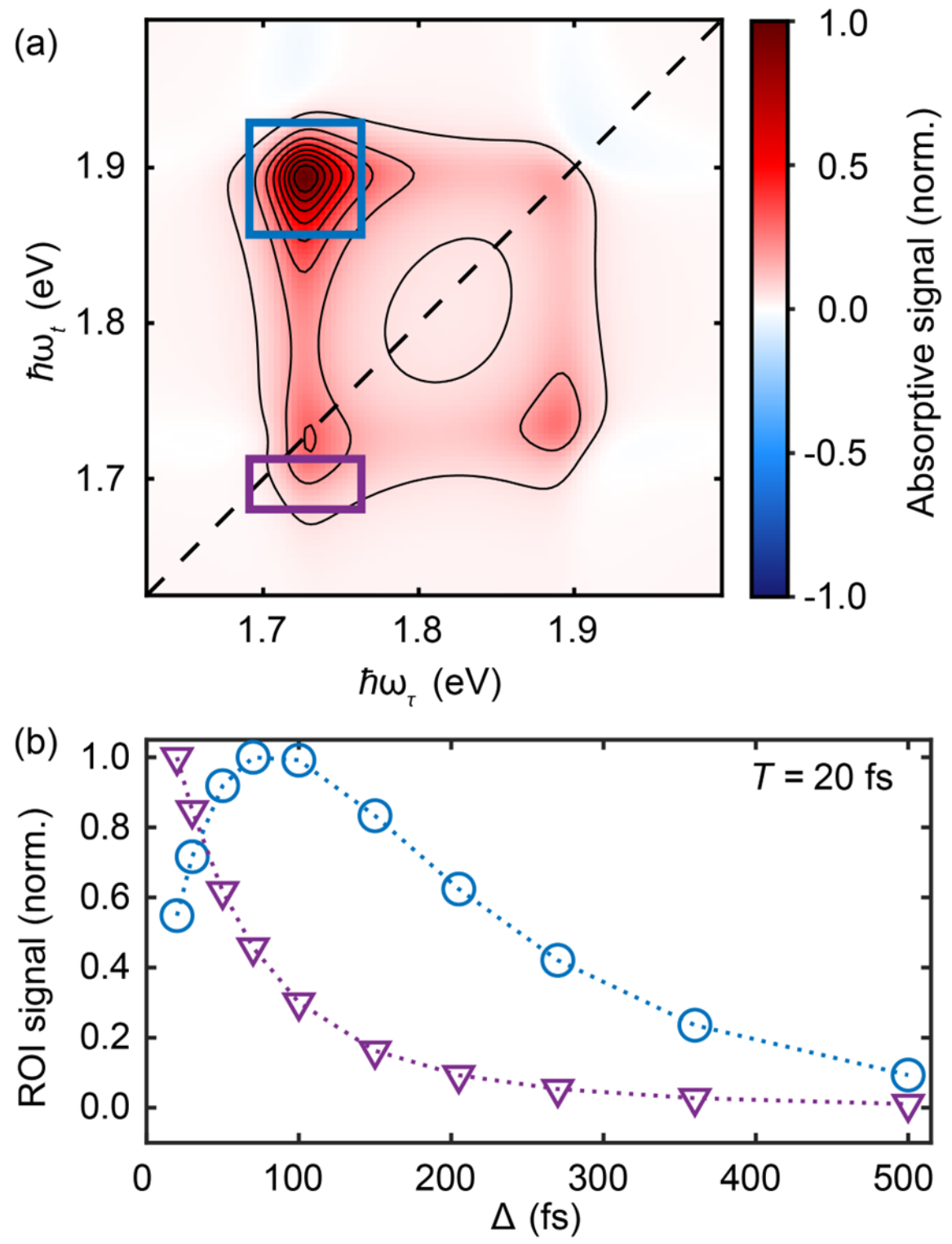

Figure 5. Simulation of kinetic-energy-filtered photoelectron-detected 2D spectroscopy. (a) 2D spectrum obtained when only the photoemission from the double-excited states $|\alpha_1\rangle$ and $|\alpha_2\rangle$ contributes to the measurement signal via kinetic-energy filtering ($T = 20$ fs and $\Delta = 20$ fs). Each square marks a region of interest (ROI). (b) Evolution of integrated ROI signals as a function of $\Delta$ at $T = 20$ fs. The dashed lines are guides to the eye.

In summary, we investigated two-dimensional electronic spectroscopy (2DES) with photoelectron detection (P-2DES) by simulating a weakly coupled molecular J-type dimer as a model system. Using an additional ionization pulse after the excitation by a four-pulse sequence enables time gating of the photoemission process and kinetic-energy filtering of the emitted photoelectrons. These two approaches can solve several challenges in action-detected 2DES, such as revealing coupling features and direct investigation of biexciton dynamics. To this end, we have expanded our open-source Matlab software package for quantum dynamics simulations, the Quantum Dynamics Toolbox (QDT),[45] to include photoelectron detection. When the time delay $\Delta$ between the ionization pulse and the last excitation pulse was kept short, spectra with features typical for coherently detected 2DES (C-2DES) were retrieved. Consequently, the coupling signatures at cross-peak positions, which are usually hidden in action-detected 2DES, were revealed. Additionally, we were able to remove the static background caused by incoherent mixing present in other action-detected techniques. This increased the signal contrast of single-exciton dynamics. When the time delay $\Delta$ was chosen long enough for efficient exciton–exciton annihilation, the spectra exhibited features typical for fluorescence-detected 2DES (F-2DES): While only single-exciton dynamics was present during the population time $T$, biexciton dynamics such as exciton–exciton annihilation could be investigated by scanning $\Delta$. By analyzing the corresponding Feynman diagrams, we showed that the contrast of different processes such as exciton–exciton annihilation and interexcitonic coherences occurring during delay times $\Delta$ and $T$, respectively, can be enhanced by a careful choice of the respective other delay time.

Our results pave the way for future experimental implementations of action-detected 2DES that suppress incoherent mixing and exploit the full potential of the method. For example, the time-gating approach could be combined with photoemission spectroscopy, angle-resolved photoemission spectroscopy (ARPES) for resolution in momentum space, or photoemission electron microscopy (PEEM) for resolution in real space. In particular, the combination of 2D spectroscopy with PEEM, so-called 2D nanoscopy, allows the acquisition of spatially resolved 2D spectra beyond the optical diffraction limit.[16] By disentangling single- and biexciton dynamics through time-gating, it becomes possible to reveal spatial defects and energy-loss channels in optoelectronic materials and surface science applications.

ASSOCIATED CONTENT

**Supporting Information**.

The following file is available free of charge:

Modeling photoemission through incoherent population transfer, Simulation parameters, Weak-coupling limit, Impact of exciton–exciton annihilation on kinetic energy resolved 2D spectrum, Ratio of ionization probabilities of states with localized character, Ratio of ionization probabilities of single-excited states, Explicit comparison of C-2DES and F-2DES with P-2DES spectra, Dependency of interexcitonic coherence signatures on ionization pulse delay $\Delta$, Dependency of annihilation signatures on energy transfer processes during $T$, Rephasing Feynman diagrams of the high-energy diagonal peak (PDF).

The data that support the findings of this study are openly available in Zenodo at https://doi.org/10.5281/zenodo.18876141/

AUTHOR INFORMATION

**Corresponding Authors**

Julian Lüttig and Tobias Brixner

Notes

The authors declare no competing financial interests.

ACKNOWLEDGMENT

T.B. acknowledges funding by the European Research Council (ERC) within Advanced Grant IMPACTS (No. 101141366). J.L. acknowledges support from the HFSP fellowship program under Grant No. LT0056/2024-C.

# Supporting Information:

# Disentangling Single- and Biexciton Dynamics with Photoelectron-Detected Two-Dimensional Electronic Spectroscopy

*Luisa Brenneis[1], Matthias Hensen[1], Julian Lüttig[1,2,*], Tobias Brixner[1,3,*]*

[1]Institut für Physikalische und Theoretische Chemie, Universität Würzburg, Am Hubland, 97074 Würzburg, Germany

[2]Department of Physics, University of Ottawa, 150 Louis-Pasteur Pvt, Church St, Ontario, ON K1N 6N5, Canada

[3]Center for Nanosystems Chemistry (CNC), Universität Würzburg, Theodor-Boveri-Weg, 97074 Würzburg, Germany

Corresponding Authors

*E-mail: julian.luettig@uni-wuerzburg.de

*E-mail: tobias.brixner@uni-wuerzburg.de

## S1. Modeling photoemission through incoherent population transfer

As described in the main text, we employ our open-source MATLAB software package, the Quantum Dynamics Toolbox (QDT),[1] to numerically propagate the time-dependent density matrix via the Lindblad master equation. The toolbox provides a suitable framework for defining the multi-level quantum system and its interaction with the pulse sequence taking into account dissipation and pure dephasing processes. Moreover, it is possible to model various nonlinear spectroscopic techniques, including coherently detected two-dimensional electronic spectroscopy (C-2DES) and fluorescence-detected two-dimensional electronic spectroscopy (F-2DES). However, to implement photoelectron emission as a detection channel in 2DES, we have extended the existing toolbox in the following way: The time-independent density matrix of the investigated quantum system is augmented by additional states that represent free-electron states $|P_i\rangle$. These free-electron states can only be populated from bound system states via an incoherent and unidirectional population transfer, whereby the transfer rate depends on the slowly varying intensity envelope of the external light field. The population transfer is incorporated through an additional Lindblad term in the Lindblad master equation, as described in the main text.

For the sake of simplicity, the continuum of free electron states is represented in our simulations by an effective state $|P_i\rangle$. Population transfer to this state is described within the Lindblad formalism through a corresponding transition rate. The index $i$ labels discrete free electron states associated with different kinetic energies. In a real system, the transition rate from a bound excited state to the continuum of free electron states depends on quantities such as the transition dipole matrix elements and density of states. In our model, these dependencies are incorporated phenomenologically through the rate constant of the respective Lindblad-type unidirectional energy transfer term describing photoemission. Accordingly, in our simulations each excited eigenstate of the system is connected to a single representative free-electron state $|P_i\rangle$ by the jump operators $\hat{\mathcal{L}}_{k'}$, providing the basis for simulating kinetic-energy-filtered spectra. As an example, we show in Figure S1 the density matrix propagation for a simplified quantum system that is driven by a four-pulse sequence and subsequently photoionized by an additional laser pulse, where $\Delta$ is the time delay between the last pulse of the four-pulse sequence and the photoionization pulse. The overall system states are depicted in Figure S1a: three bound states $|g\rangle$, $|\varepsilon\rangle$, and $|\alpha\rangle$, and two free-electron states $|P_\varepsilon\rangle$ and $|P_\alpha\rangle$, which are related to the photoionization of $|\varepsilon\rangle$ and $|\alpha\rangle$, respectively. Figures S1b to f show the diagonal elements of the time-dependent density matrix. Here, the evolution is depicted as a function of the simulation time $t'$ for $\phi_1 = \phi_2 = \phi_3 = \phi_4$, $\tau = t = 224$ fs and $T = \Delta = 500$ fs. Each pulse of the four-pulse sequence as well as the ionization pulse exhibits a full width at half maximum (FWHM) of 10 fs.

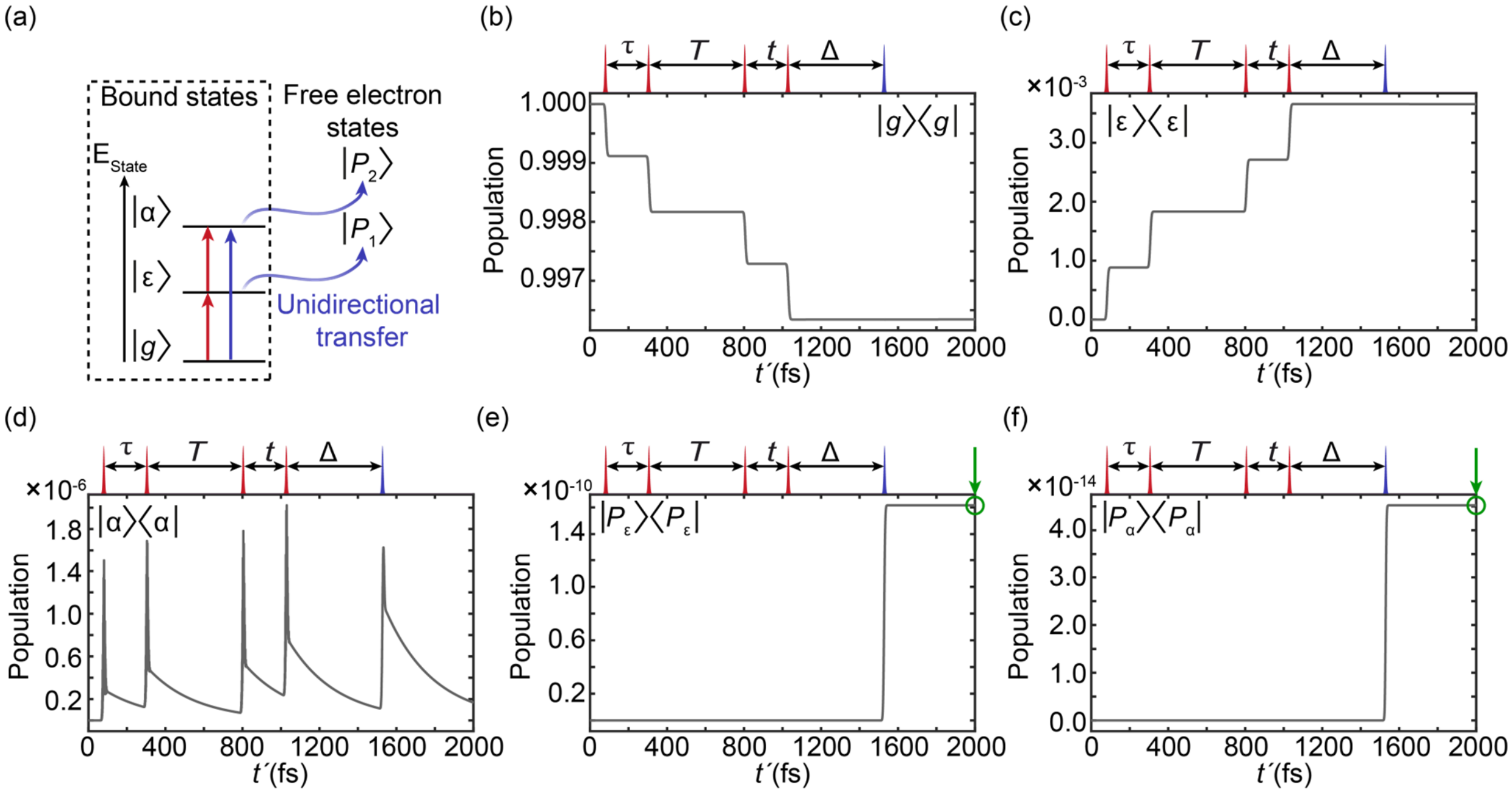

Figure S1. Evolution of real part of the density matrix diagonal elements for a simplified model system. (a) Simplified model system with three bound states ($|g\rangle$, $|\varepsilon\rangle$, and $|\alpha\rangle$) and two free-electron states ($|P_\varepsilon\rangle$ and $|P_\alpha\rangle$). The free-electron states are solely connected via an unidirectional population transfer with their corresponding bound state. (b–g) Time-dependent density matrix elements for a specific set of pulse sequence parameters ($\phi_1 = \phi_2 = \phi_3 = \phi_4$, $\tau = t = 224$ fs, and $T = \Delta = 500$ fs): (b) ground state, (c) single-excited state, (d) double-exited state, (e) first free-electron state $|P_\varepsilon\rangle$, and (f) second free-electron state $|P_\alpha\rangle$. Green arrows point to the single point of time per free electron state used to calculate the total photoemission yield.

Before the first pulse interaction ($t' < 80$ fs), only the ground state $|g\rangle$ is populated (Figure S1b) and all other density matrix elements are zero. During the interactions with each pulse of the phase-cycled four-pulse sequence, the states $|\varepsilon\rangle$ and $|\alpha\rangle$ get populated while the ground state $|g\rangle$ is depopulated. During the interval between the light–matter interactions, the excited-state populations can dissipate in this simplified system via a relaxation from $|\alpha\rangle$ to $|\varepsilon\rangle$ with a rate constant of $1/250$ fs$^{-1}$. As we assume that no multi-photon ionization occurs, the free-electron states (Figure S1e and f) are only populated through the unidirectional energy transfer, whose rate constant depends on the intensity envelope of the ionization pulse (blue pulse). Note that the neglection of multi-photon absorption is not required for our approach to model photoelectron detection in 2D spectroscopy. Including multi-photon ionization that is triggered by the red multi-pulse sequence would just lead to an additional $\Delta$-independent background. This background could be removed in an experiment by an appropriate chopping scheme.

The population of free-electron states is free from relaxation processes and is therefore stored until the end of the simulation time $t'$. In this way, we account for the collection of the emitted photoelectrons by the detector. Note that only a small part of the excited-state population is transferred into the free-electron states because the amplitude of the ionization pulse is kept in a perturbative regime. Further note that the population of the state $|\alpha\rangle$ increases when the ionization pulse interacts with the bound system states as linear absorption from the ground state $|g\rangle$ occurs. The total photoemission yield $Y$ is calculated by summing over the final population value of all free-electron states (Figure S1e and f, green arrows). For kinetic-energy-filtered spectra, we select specific free-electron state populations.

To simulate nonlinear spectroscopy experiments, the density matrix propagation is calculated for each pulse sequence as a function of the time delays $\tau$, $T$, $t$, and $\Delta$ as well as for the interpulse phase differences $\phi_2-\phi_1$, $\phi_3-\phi_1$, and $\phi_4-\phi_1$. Notably, the calculated density matrix dynamics inherently include contributions from all orders of nonlinearity, which becomes evident when comparing the absolute amplitudes of Figure S1c and d. The population of state $|\varepsilon\rangle$ contains the desired fourth-order response but also lower-order contributions, such as single-photon absorption (second-order processes). These undesired contributions are removed during post-processing via phase cycling.[2] The spikes observed in Figure S1d originate from coherent interference between pathways during the pulse interaction intervals and exhibit a strong dependence on the dephasing times of the involved states as well as on the pulse duration.

## S2. Simulation parameters

In the main text, we consider a purely excitonic heterodimer as a model system. The transition dipole moments and energies are based on a squaraine heterodimer.[3] The dissipation and dephasing rates are chosen to allow clear demonstration of the temporal evolution of specific features in the 2D spectra as a function of the delay times $T$ or $\Delta$, and to unambiguously separate individual spectral contributions, while remaining within a physically reasonable regime typical of weakly coupled molecular systems. Figure S2 summarizes the system properties in the site (Figure S2a) and exciton basis (Figure S2b). All system-dependent simulation parameters are listed in Table S1. Due to weak coupling, the exciton eigenstates $|\alpha_1\rangle$ and $|\alpha_2\rangle$ are strongly localized on the individual sites $|f_A\rangle$ and $|f_B\rangle$ of the site basis, with weights of 99.5 % and 98.3 %, respectively, as given by the squared magnitudes of the transformation coefficients. In contrast, the exciton eigenstate $|\alpha_3\rangle$ has a strong contribution of the delocalized state $|e_A e_B\rangle$, with a weight of 97.9 %.

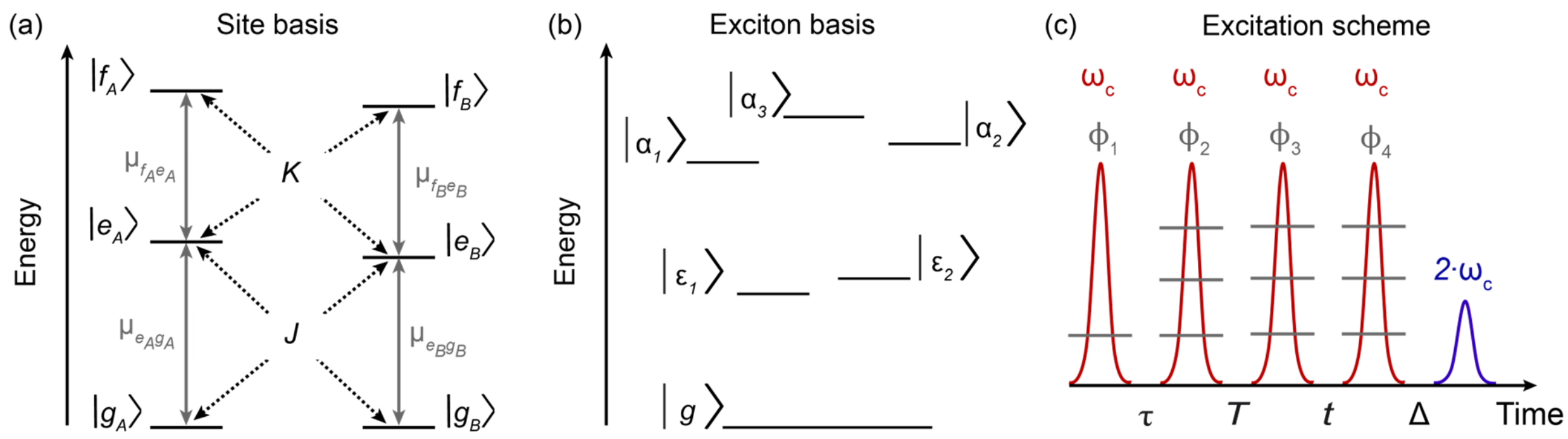

Figure S2. Schematic illustration of simulation parameters. (a) System states represented in the site basis. (b) System states represented in the exciton basis. (c) Excitation scheme.

For each $T$ and $\Delta$ step, the simulation was performed on a temporal grid with a grid step size of $\Delta t' = 0.1$ fs and the first pulse interaction was delayed by 80 fs with respect to the starting point at $t' = 0$ fs. The excitation scheme is schematically illustrated in Figure S2c. The pulse duration of each pulse of the four-pulse sequence (Figure S2c, red pulses) as well as of the ionization pulse (Figure S2c, blue pulse) was set to 10 fs (FWHM). To subtract lower-order contributions, a $1 \times 3 \times 3 \times 3$ phase-cycling scheme was implemented for the four-pulse sequence. The time delays $\tau$ and $t$ were scanned from 0 fs to 224 fs in 33 steps in a fully rotating frame with a central frequency of 437 THz ($\hbar\omega_c = 1.81$ eV). Fourier transformation along $\tau$ and $t$ was performed after 5-fold zero padding along each axis. The central photon energy of the ionization pulse was set to twice of the four-pulse sequence pulses, i.e., $\hbar\omega_{c,\mathrm{ion}} = 2 \times 1.81\ \mathrm{eV} = 3.62\ \mathrm{eV}$. The amplitude of the ionization pulse was set at a fraction of 0.1 of the amplitude of each temporally separated pulse of the four-pulse-sequence to avoid significant ionization from the ground state via multi-photon absorption of the ionization pulse.

Table S1. Simulation parameters for the system presented in the main text.

| Quantity | | Value |
|---|---|---|
| Transition | $E_{e_A g_A}$ | 1.890 |
| | $E_{f_A e_A}$ | 1.612 |
| | $E_{e_B g_B}$ | 1.730 |
| | $E_{f_B e_B}$ | 1.612 |
| Transition dipole moments | $\mu_{e_A g_A}$ | 1.920 |
| | $\mu_{f_A e_A}$ | 1.267 |
| | $\mu_{e_B g_B}$ | 1.410 |
| | $\mu_{f_B e_B}$ | 0.931 |
| Coupling | $J$ | 29.7 |
| | $K$ | 23.8 |
| Dissipation | $T_{\mathrm{Dis}}$ | 300 fs |
| | $T_{\mathrm{Dis}}$ | 300 fs |
| | $T_{\mathrm{Dis}}$ | 50 fs |
| | $T_{\mathrm{Dis}}$ | 50 fs |
| | $T_{\mathrm{Dis}}$ | 250 fs |
| Pure dephasing | all | 200 fs |

For the simulation of the fluorescence-detected spectrum (Section S7), the same parameters as described above were used. However, the amplitude of the ionization pulse was set to zero, and the maximum simulation time $t'$ was chosen such that all populations relaxed to the lowest excited state before the simulation ended. The signal was then calculated as the sum over the integrated bound excited-state populations of the density matrix.

For the simulation of the C-2DES spectrum (Section S7) the time delay $\tau$ between the two pump pulses was scanned from 0 fs to 100 fs in 301 steps in a partially rotating frame ($\gamma = 0.4$) with a central frequency of 437 THz ($\hbar\omega_c = 1.81$ eV). The probe interaction was time-delayed with respect to the second pump interaction by $T = 20$ fs and was $1 \times 2$ phase-cycled. The amplitude of the probe pulse was set to 30 % of the amplitude of one of the temporally separated pump pulses. The simulation step size was set to $\Delta t' = 0.2$ fs. The signal was calculated as the sum of the off-diagonal elements of the density matrix from the final pulse interaction until 240 fs after the final pulse interaction and multiplied by the corresponding transition dipole moments, as described by Malý et al.[4] Fourier transformation of this time-dependent signal provided the $\omega_t$-dependent signal for each $\tau$ and $T$ step and the absorptive signal was calculated subsequently.

## S3. Weak-coupling limit

The investigation of annihilation dynamics in weakly coupled systems is particularly relevant due to its connection to exciton diffusion time constants.[5] In strongly coupled systems, by comparison, annihilation processes typically occur on much shorter timescales due to enhanced exciton delocalization and transport. In this regime, the dynamics can no longer be fully characterized in terms of simple rate equations and well-defined annihilation time constants, even if the experimental temporal resolution remains sufficient to resolve the corresponding ultrafast processes. The applicability of our conclusions to more strongly coupled systems depends on whether the EEA mechanism can still be described within a sequential two-step picture. This remains, in principle, valid also in the case of delocalized excitons, if the biexciton states can be represented predominantly as superpositions of the single-exciton states. However, if the biexciton states mix significantly with the double-excited states of the monomers, the two sequential steps of EEA (exciton fusion and subsequent decay into the single-exciton manifold) can no longer be treated separately. Instead, the system is described by a manifold of delocalized eigenstates. Consequently, the ionization probability becomes strongly influenced by hybridization with these double-excited states, and whether EEA signatures can still be isolated becomes highly system dependent.

In the following, we relate this discussion to our model system. The higher-excited states $|\alpha_1\rangle$, $|\alpha_2\rangle$ and $|\alpha_3\rangle$ are linear combinations of double-excited site states and the biexciton state $|e_A e_B\rangle$ . In particular,

$$|\alpha_3\rangle = c_1|e_A e_B\rangle + c_2|g_A f_B\rangle + c_3|f_A g_B\rangle, \tag{S1}$$

where, in the weak-coupling limit, $c_1 \gg c_2$ and $c_1 \gg c_3$. Consequently, $|\alpha_3\rangle$ is predominantly biexcitonic in character and the time-dependent population $|\alpha_3\rangle\langle\alpha_3|$ is governed by the biexciton dynamics. The corresponding relative yield can therefore be approximated as $\Gamma_{\alpha_3} \approx \Gamma_{e_A e_B}$. Since ionization can occur independently from both singly excited sites, the biexciton ionization yield is $\Gamma_{e_A e_B} = \Gamma_{e_A} + \Gamma_{e_B}$. By the same argument, the lower excited states are predominantly localized, such that $\Gamma_{\varepsilon_1} \approx \Gamma_{e_B}$ and $\Gamma_{\varepsilon_2} \approx \Gamma_{e_A}$. Consequently, the factor of the state with biexciton character $\Gamma_{\alpha_3}$is larger than the factor of the single-excited states $\Gamma_{\varepsilon_1}$ or $\Gamma_{\varepsilon_2}$. The relation $\Gamma_{\alpha_3} > \Gamma_{\varepsilon_1}$ leads to a $\Delta$-dependent evolution of ESA2 pathways which is governed exclusively by biexciton relaxation.

In a more strongly coupled system, $\Gamma_{\alpha_3}$ also depends on the ionization probability of the double-excited states of the monomers. Specifically, whether $\Gamma_{\alpha_3} > \Gamma_{\varepsilon_1}$ is determined by the ratios $\Gamma_{f_A}/\Gamma_{e_A}$ and $\Gamma_{f_B}/\Gamma_{e_A}$, which are system dependent. Additionally, the evolution of $|\alpha_3\rangle\langle\alpha_3|$ can no longer be described in the two-step EEA model.

In the present work, the photoemission process is modeled using local ionization channels of the individual sites, which is well justified in the weak-coupling limit where the excited states remain predominantly localized. Beyond this limit, the approximation breaks down due to the increasing delocalization of the electronic states. A more advanced treatment of the photoemission process would then be required, which is beyond the scope of this work.

## S4. Impact of exciton–exciton annihilation on the kinetic-energy-resolved 2D spectrum

Figure S3 presents a kinetic-energy-filtered 2D spectrum that was obtained in a similar way as the 2D spectrum in Figure 5a, with the only difference being an artificially suppressed exciton–exciton annihilation (EEA) rate. Consequently, relaxation from $|\alpha_3\rangle$ is prevented and only signal contributions summarized by the ESA2 Feynman diagram in the top right of Figure S3 contribute. Each subscript $l, m,$ and $n$ in the diagram may assume either the value 1 or 2. While only a nonrephasing diagram is shown, a corresponding rephasing diagram also exists, and the following discussion applies analogously to the rephasing contributions. The square-shaped regions of interest (ROI) in the 2D spectrum are identical to those in Figures 3a and 5a and are provided for better orientation. The main signal amplitude in Figure S3 is located slightly below the diagonal at $\hbar\omega_\tau = 1.72$ eV and $\hbar\omega_t = 1.71$ eV, corresponding to the energy gaps between $|g\rangle$ and $|\varepsilon_1\rangle$, and between $|\varepsilon_1\rangle$ and $|\alpha_2\rangle$, respectively. The comparison with Figure 5a reveals that at $\hbar\omega_\tau = 1.72$ eV and $\hbar\omega_t = 1.89$ eV (blue ROI), where the maximal amplitude is located when notable EEA is present, the signal is weak and no cross peak appears in the case without EEA. We therefore conclude that the peak in Figure 5a at $\hbar\omega_\tau = 1.72$ eV, $\hbar\omega_t = 1.89$ eV (blue ROI) arises predominantly from the initial population of the $|\alpha_3\rangle$ state and subsequent relaxation. Therefore, in Figure 5b, we can observe the rise and subsequent decay of the cross peak (blue ROI), where the decay is associated with the relaxation from $|\alpha_3\rangle$, while slightly below the diagonal (purple ROI) the relaxation from $|\alpha_2\rangle$ dominates.

Due to kinetic-energy filtering, only ESA2 diagrams contribute to the 2D spectrum. When the total photoemission yield is considered, these contributions destructively interfere with the corresponding ESA pathways. Especially, each ESA2 pathway ending in a population of $|\alpha_1\rangle$ or $|\alpha_2\rangle$ (Figure S3, top) cancels with the corresponding ESA pathway (Figure S3, bottom) as both contain not only the same transition dipole moments but also the same $\Gamma_j$ factor of the ionization process as, in our model, $\Gamma_{\alpha_n} = \Gamma_{\varepsilon_m}$. Therefore, none of the features of the spectrum in Figure S3 appears in the 2D spectrum obtained by the total photoemission yield.

However, in systems where $\Gamma_{\alpha_n} \neq \Gamma_{\varepsilon_m}$, a residual signal contribution would remain. If $\Gamma_{\alpha_n} > \Gamma_{\varepsilon_m}$, the ESA2 contribution dominates, resulting in an additional positive signal in the kinetic-energy-integrated 2D spectrum, whereas $\Gamma_{\alpha_n} < \Gamma_{\varepsilon_m}$ leads to an additional negative signal.

In C-2DES, such ESA signals are also present exhibiting a positive sign. For the present model system, Figure S3 indicates that these additional ESA contributions would predominantly overlap with the low-energy diagonal peak located at $\hbar\omega_\tau = \hbar\omega_t = 1.72$ eV.

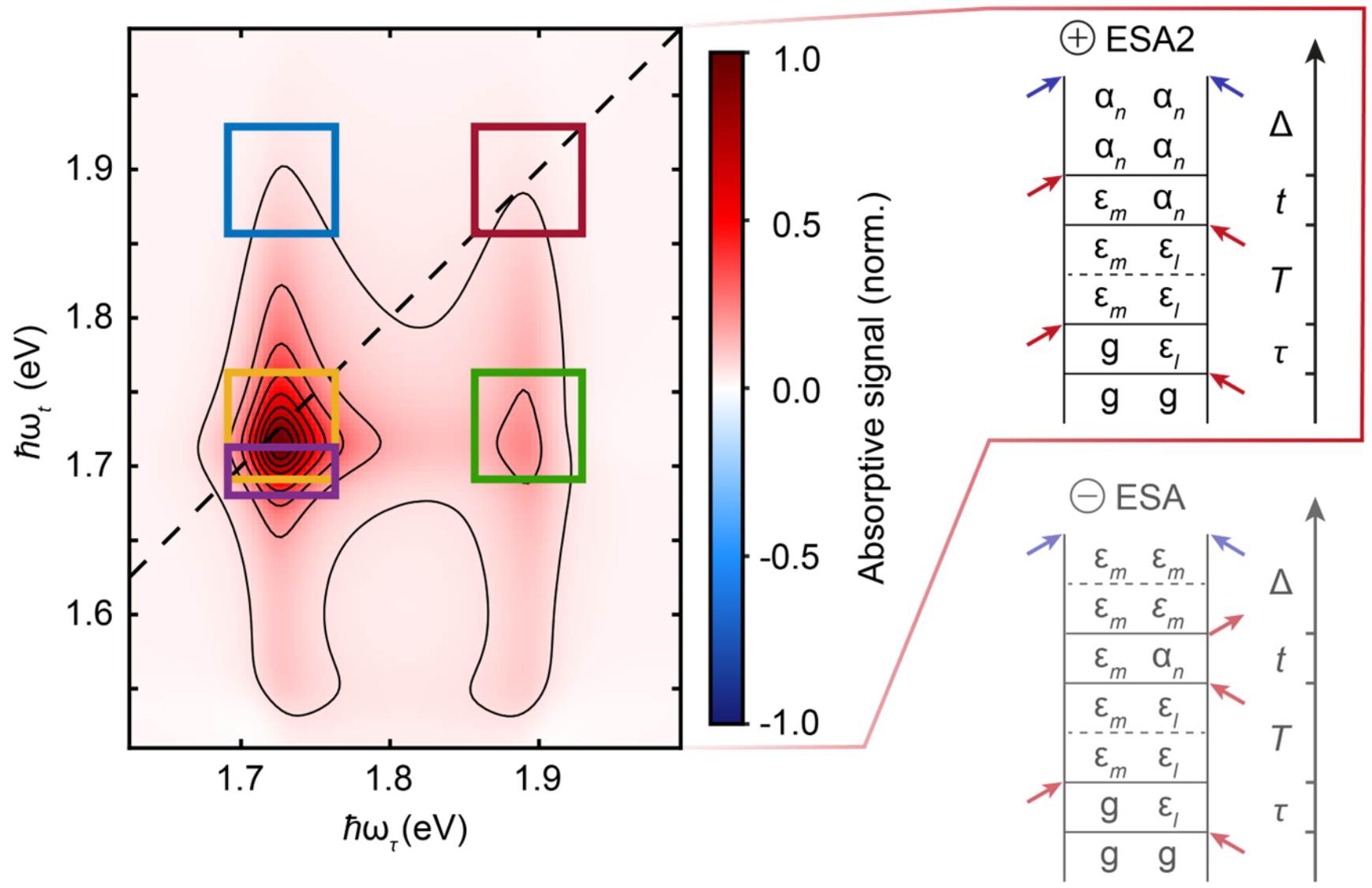

Figure S3. Simulation of a kinetic-energy-filtered photoelectron-detected 2D spectrum without exciton–exciton annihilation. Each square marks the same region of interest (ROI) as in Figure 3a and Figure 5a of the main text. Only ESA2 pathways, as illustrated on the top right, contribute to the spectrum, where all subscripts can either represent 1 or 2. In a kinetic-energy-integrated 2D spectrum, the ESA2-type contributions destructively interfere with negative ESA pathways shown on the bottom right. While only nonrephasing diagrams are presented here, corresponding rephasing diagrams exist, too. Time ordering of all pulses is assumed with respect to the sequence shown in Figure 1b. Red arrows represent interactions with the phase-cycled four-pulse sequence while blue arrows indicate interactions with the ionization pulse.

## S5. Ratio of ionization probabilities of states with localized character

In the main text, we considered a special case in which all ionization probability factors of states with strong localized character are identical, i.e. $\gamma_{\alpha_2} = \gamma_{\alpha_1} = \gamma_{\varepsilon_2} = \gamma_{\varepsilon_1}$ and $\gamma_{\varepsilon_1} := 1$. Here we extend this analysis to a more general scenario where the ionization probabilities of states with predominantly one double-excited site ($\gamma_{\alpha_2}, \gamma_{\alpha_1}$) differ from those of the single-excited states ($\gamma_{\varepsilon_2}$, $\gamma_{\varepsilon_1}$). Since these ratios depend on the correlated cationic state manifolds and the individual transition rates, they are inherently system specific. We therefore systematically vary these ratios to demonstrate that the biexciton relaxation dynamics can still be extracted even when $\gamma_{\alpha_1} = \gamma_{\alpha_2} \neq \gamma_{\varepsilon_1} = \gamma_{\varepsilon_2}$.

We vary the ratio $\gamma_{\alpha_x}/\gamma_{\varepsilon_1}$ with $x \in \{1,2\}$ while keeping the other parameters fixed to those in the main text ($\gamma_{\varepsilon_2} = \gamma_{\varepsilon_1}, \gamma_{\alpha_3} = \gamma_{\varepsilon_2} + \gamma_{\varepsilon_1} = 2\,\gamma_{\varepsilon_1}, \gamma_{\alpha_2} = \gamma_{\alpha_1}$). We start with the analysis at the above-diagonal cross-peak position ($\hbar\omega_\tau = 1.72$ eV, $\hbar\omega_t = 1.89$ eV), i.e., the spectral region exhibiting the strongest EEA signature in the main text. Figure S4a shows representative dynamics of the integrated ROI signal for four ratios $\gamma_{\alpha_x}/\gamma_{\varepsilon_1}$. All data are normalized to the absolute global maximum across all ROIs defined in Figure 3a. The solid lines correspond to the biexponential fits using

$$Y(\Delta) = (F_{A'} + F_{B'}) \cdot e^{\frac{-2\Delta}{\tau_1}} - F_{B'} \cdot e^{\frac{-\Delta}{\tau_2}} + c \qquad \text{(S2)}$$

with the amplitudes $F_{A'}$ and $F_{B'}$, which depend on the population of all higher-excited states $|\alpha_1\rangle$, $|\alpha_2\rangle$ and $|\alpha_3\rangle$ after the fourth pulse interaction at the investigated ROI and the corresponding ionization probability factors. The parameters $\tau_1$ and $\tau_2$ represent the time constants of the relaxation processes of these states. In our model $|\alpha_3\rangle\langle\alpha_3|$ decays exponentially with $\tau_1$ into $|\alpha_2\rangle\langle\alpha_2|$ and $|\alpha_1\rangle\langle\alpha_1|$. The populations $|\alpha_2\rangle\langle\alpha_2|$ and $|\alpha_1\rangle\langle\alpha_1|$ rise with the time constant $\tau_1$ (due to the relaxation from $|\alpha_3\rangle\langle\alpha_3|$) and decay with $\tau_2$ corresponding to the relaxation into the single-exciton manifold. If $|\alpha_2\rangle\langle\alpha_2|$ and $|\alpha_1\rangle\langle\alpha_1|$ are not probed by the four-pulse sequence, $F_{A'}$ only depends on the population $|\alpha_3\rangle\langle\alpha_3|$ at $\Delta = 0$ fs and $\gamma_{\alpha_3}$.

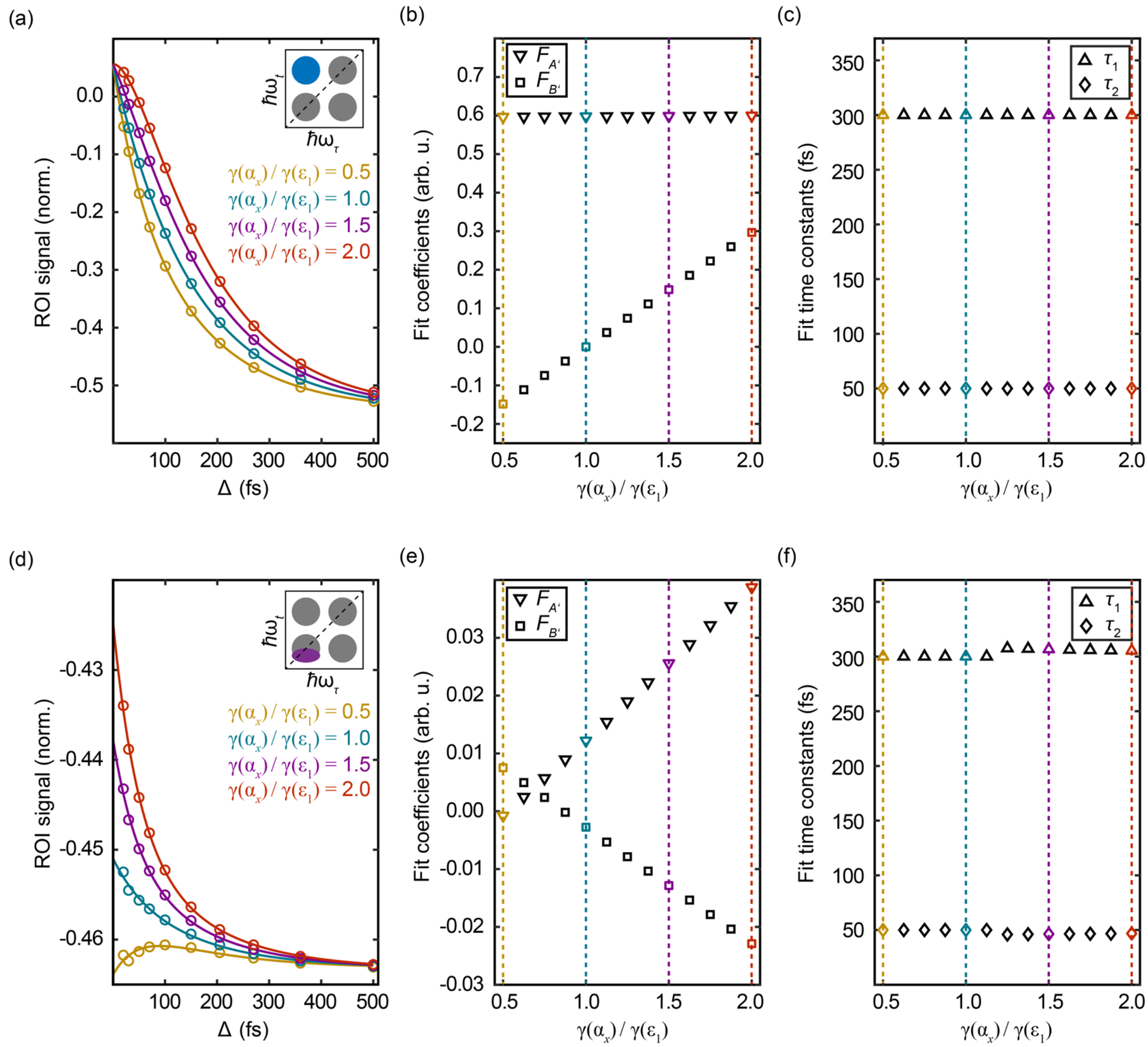

Figure S4. Dependency of the time-dependent signal on the ratio $\gamma_{\alpha_x}/\gamma_{\varepsilon_1}$ with $x \in \{1,2\}$ as a function of delay time $\Delta$. (a) Representative time traces of the above-diagonal cross-peak ROI at ($\hbar\omega_\tau = 1.72$ eV, $\hbar\omega_t = 1.89$ eV) for $T = 20$ fs. Solid lines show exponential fits according to Equation S2. The inset illustrates the spectral position of the ROI in the 2D spectra, as defined in Figure 3a. (b–c) Extracted fit amplitudes (b) and time constants (c) obtained from Equation S2 as a function of $\gamma_{\varepsilon_2}/\gamma_{\varepsilon_1}$ for the above-diagonal cross-peak ROI. Colored markers and vertical lines correspond to the fit results of the traces shown in (a). (d) Representative time traces of the ROI centered at $\hbar\omega_\tau = 1.72$ eV, $\hbar\omega_t = 1.71$ eV for $T = 20$ fs. Solid lines show exponential fits according to Equation S2. The inset indicates the ROI position in the 2D spectrum. (e–f) Extracted amplitudes (e) and time constants (f) obtained from fits with Equation S2 as a function of $\gamma_{\varepsilon_2}/\gamma_{\varepsilon_1}$ for the ROI signal of (d). Colored markers and vertical lines indicate the fit results of the traces shown in (d).

All four depicted time traces overlap at $\Delta = 0$ fs and at large delay times $\Delta$, and show different dynamics in between. The excitations from the single-exciton manifold ($|\varepsilon_x\rangle$) to the localized excited states $|\alpha_x\rangle$ do not contribute to the above-diagonal cross peak. Therefore, the $\gamma_{\alpha_x}/\gamma_{\varepsilon_1}$ ratio has no impact on the yield at $\Delta = 0$ fs. All fits converge to the same signal value at large delay times $\Delta$ since the ratio $\gamma_{\alpha_x}/\gamma_{\varepsilon_1}$ does not affect the signal once all populations have relaxed into $|\varepsilon_1\rangle$. Figures S4b and S4c summarize the coefficients of the fits for $\gamma_{\alpha_x}/\gamma_{\varepsilon_1}$ ratios ranging from 0.5 to 1.5. The extracted time constants $\tau_1$ and $\tau_2$ (Figure S4c) are in excellent agreement with $T_{Dis}(\alpha_3 \rightarrow \alpha_x) = 300$ fs and $T_{Dis}(\alpha_x \rightarrow \varepsilon_x) = 50$ fs, respectively. These timescales reflect the EEA kinetics: population transfer from the biexcitonic state with predominantly two singly excited sites to predominantly double-excited localized states ($T_{Dis}(\alpha_3 \rightarrow \alpha_x) = 300$ fs), followed by rapid relaxation into the single-exciton manifold ($T_{Dis}(\alpha_x \rightarrow \varepsilon_x) = 50$ fs). Notably, in the case discussed in the main ($\gamma_{\alpha_x}/\gamma_{\varepsilon_1} = 1$), the $\Delta$-dependent evolution only depends on the first transition step, corresponding to $T_{Dis}(\alpha_3 \rightarrow \alpha_x)$. Consequently, the signal relaxes with a time constant of $\tau_1/2 = 150$ fs (see Equation S2). In this regime, the uncertainty of the extracted time constant $\tau_2$ is significantly larger than for other $\gamma_{\alpha_x}/\gamma_{\varepsilon_1}$ ratios. However, this does not affect the overall quality of the fit (Figure S4a, turquoise line), since $F_{B'} = 0$ (see Figure S4b). For other cases, both annihilation steps contribute to the time trace dynamics: the exciton fusion $T_{Dis}(\alpha_3 \rightarrow \alpha_x)$ as well as the fast decay $T_{Dis}(\alpha_x \rightarrow \varepsilon_x)$. If the internal conversion occurs instantaneously on the time scale of the exciton fusion, the annihilation dynamics can be described by respective rate expressions.[6] Similarly, if the time constants $\tau_1 \gg \tau_2$, Equation S2 effectively reduces to a single exponential decay at longer delay times. Figure S4b further shows that only $F_{B'}$ varies with the ratio $\gamma_{\alpha_x}/\gamma_{\varepsilon_1}$, while $F_{A'}$ remains constant. This behavior indicates that the states $|\alpha_1\rangle$ and $|\alpha_2\rangle$ are not directly populated at $\Delta = 0$ fs at this spectral position, consistent with the discussion in Section S4. Consequently, the initial amplitude at $\Delta = 0$ fs is unaffected by the $\gamma_{\alpha_x}/\gamma_{\varepsilon_1}$ ratio, and the $\Delta$-dependent evolution arise exclusively from biexciton relaxation into the single-exciton manifold, i.e., EEA.

At other spectral positions, corresponding to transitions $\varepsilon_x \rightarrow \alpha_x$, the states $|\alpha_x\rangle$ states can be directly populated. If these transitions lie within the laser bandwidth, they give rise to additional ESA features at early delay times $\Delta$ when $\gamma_{\alpha_x} \neq \gamma_{\varepsilon_1}$ (see Section S4), which decay fast into the

single-exciton manifold as a function of the delay $\Delta$. In our specific example, most of these transitions lie outside the laser spectrum, except for the transition between $|\varepsilon_1\rangle$ and $|\alpha_2\rangle$ at $\hbar\omega_t =$ 1.71 eV, which overlaps with the low-energy diagonal peak at $\hbar\omega_t = 1.72$ eV. Representative time traces at this position are shown in Figure S4d. Despite the partial spectral overlap, the traces exhibit a pronounced fast decay for $\gamma_{\alpha_x}/\gamma_{\varepsilon_1} > 1$. Fits using Equation S2 confirm that the dynamics are dominated by the fast timescale $\tau_2 \approx 50$ fs (Figures S4f) as $|(F_{A'} + F_{B'})| \ll |F_{B'}|$ (Figures S4e) because $F_{A'}$ and $F_{B'}$ have different signs.

## S6. Ratio of ionization probabilities of single-excited states

In this section, we analyze the case with different ionization probabilities of the two single-excited states ($\gamma_{\varepsilon_2} \neq \gamma_{\varepsilon_1}$), which is particularly relevant for heterodimer systems with strongly differing monomer contributions. In the main text we considered a special case with the factors $\gamma_{\alpha_1} = \gamma_{\alpha_2} = \gamma_{\varepsilon_1}$, $\gamma_{\varepsilon_2} = \gamma_{\varepsilon_1}$, $\gamma_{\alpha_3} = \gamma_{\varepsilon_2} + \gamma_{\varepsilon_1} = 2\,\gamma_{\varepsilon_1}$. Here, we vary the ratio $\gamma_{\varepsilon_2}/\gamma_{\varepsilon_1}$ while keeping the parameters of the double-excited states ($\gamma_{\alpha_1}$, $\gamma_{\alpha_2}$) constant. Throughout this analysis, we fix $\gamma_{\varepsilon_1} = 1$ and vary $\gamma_{\varepsilon_2}$, which consequently also changes $\gamma_{\alpha_3} = \gamma_{\varepsilon_2} + \gamma_{\varepsilon_1}$.

As discussed in the main text, in the general case where $\gamma_{\varepsilon_2} \neq \gamma_{\varepsilon_1}$, both EEA and single-exciton energy transfer alter the photoelectron signal as a function of $\Delta$. Nevertheless, both contributions can still be disentangled using the information about single-exciton dynamics provided by the $T$-dependent data. Figure S5 shows the temporal evolution of the high-energy diagonal peak (Figure S5a) and the below-diagonal cross peak (Figure S5b) as a function of $T$ for three exemplary ratios $\gamma_{\varepsilon_2}/\gamma_{\varepsilon_1}$. The solid lines are fits using

$$Y(T) = A \cdot e^{\frac{-T}{\tau_3}} + c, \tag{S3}$$

where a single exponential decay describes the relaxation dynamics in the single-exciton manifold. As in the main text, data points for $T < 150$ fs were excluded in order to avoid contributions from interexcitonic coherences and pulse overlap. This analysis was repeated for a series of $\gamma_{\varepsilon_2}/\gamma_{\varepsilon_1}$ ratios (Figure S5c). The extracted time constants for both ROI positions are shown in Figure S5c. In all cases, the fitted time constants $\tau_3$ agree with the single-exciton energy transfer time associated with the transition from $|\varepsilon_2\rangle$ to $|\varepsilon_1\rangle$. This information can subsequently be used to isolate the EEA dynamics from the $\Delta$-dependent time evolution. This is demonstrated in the following.

As described in the main text, the $\Delta$-dependent time evolution is sensitive to dynamics between populations of states exhibiting a different factor associated with the ionization probability. In the present scenario, $\gamma_{\varepsilon_2}/\gamma_{\varepsilon_1} \neq 1$ and $\gamma_{\alpha_3}/\gamma_{\varepsilon_1} \neq 1$. Therefore, the population changes of $|\alpha_3\rangle\langle\alpha_3|$ and $|\varepsilon_2\rangle\langle\varepsilon_2|$ change the signal as a function of $\Delta$. In our model, the biexcitonic state $|\alpha_3\rangle\langle\alpha_3|$ shows a monoexponential decay due to EEA into $|\alpha_x\rangle\langle\alpha_x|$ with $x \in \{1,2\}$, associated with $\tau_1$. The population $|\varepsilon_2\rangle\langle\varepsilon_2|$ relaxes monoexponentially into $|\varepsilon_1\rangle\langle\varepsilon_1|$, associated with the time constant $\tau_3$. This time constant is already known from the $T$-dependent traces. Due to EEA, the population $|\varepsilon_2\rangle\langle\varepsilon_2|$ rises with the second time constant of the annihilation process ($\tau_2$). We fit the integrated ROI data of the above-diagonal cross peak using

$$Y(\Delta) = F_A \cdot e^{\frac{-2\Delta}{\tau_1}} + F_B \cdot e^{\frac{-\Delta}{\tau_2}} + F_C \cdot e^{\frac{-\Delta}{\tau_3}} + c, \qquad \text{(S4)}$$

with $\tau_3$ fixed to 250 fs. Equation S4 exhibits three exponential functions with the time constants $\tau_1$, $\tau_2$, and $\tau_3$. The amplitudes $F_A$, $F_B$, and $F_C$ depend on the initially excited populations at $\Delta = 0$ fs. The corresponding results and exemplary time traces of the above-diagonal cross-peak position are shown in Figure S6. The extracted time constants (Figure 6c) are in excellent agreement with $T_{Dis}(\alpha_3 \rightarrow \alpha_x) = 300$ fs and $T_{Dis}(\alpha_x \rightarrow \varepsilon_x) = 50$ fs. These timescales correspond to the EEA dynamics: The population transfer from the biexcitonic state, characterized by (predominantly) two singly excited sites, to double-excited localized states (300 fs), followed by rapid relaxation into the single-exciton manifold within 50 fs.

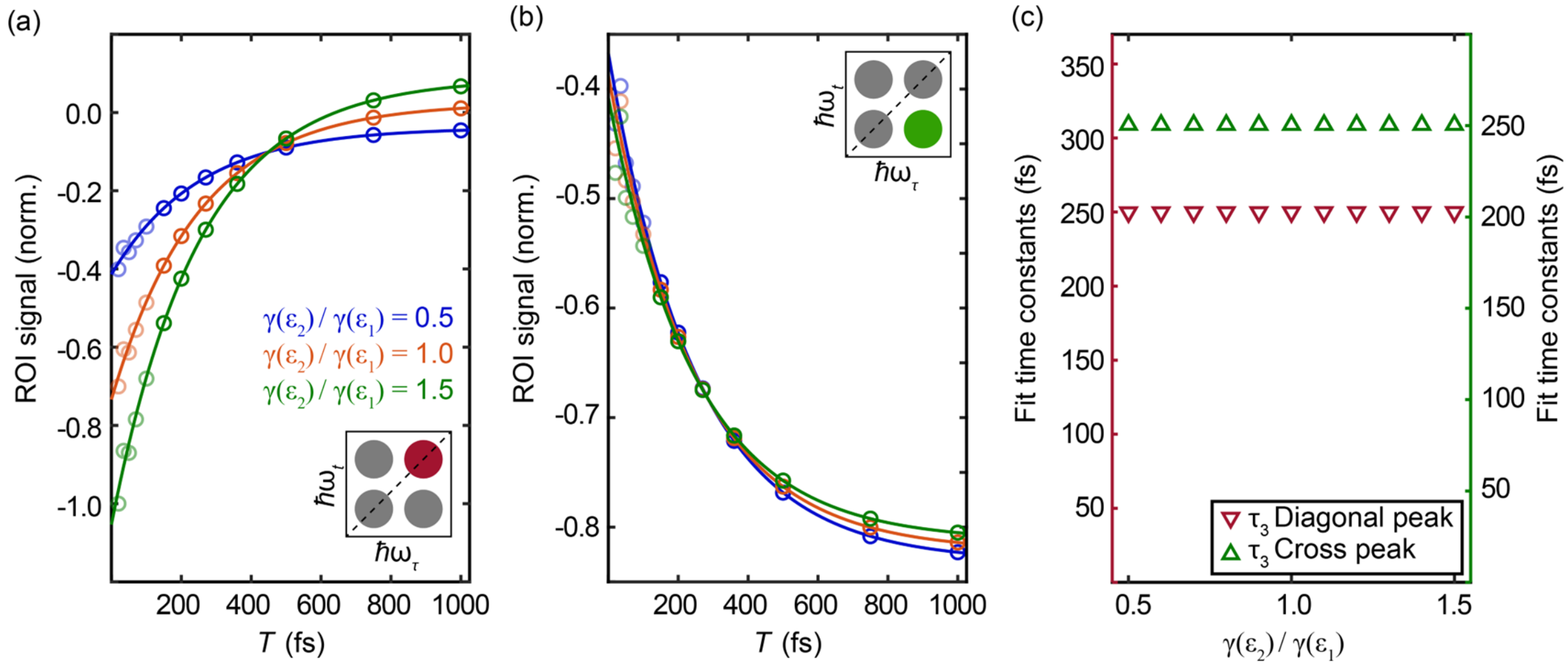

Figure S5. Dependency of the time-dependent cross-peak ROI signal as a function of the delay time $T$ on the ratio $\gamma_{\varepsilon_2}/\gamma_{\varepsilon_1}$. (a) High-energy diagonal peak ROI ($\hbar\omega_\tau = 1.89$ eV, $\hbar\omega_t = 1.89$ eV) signal for $\Delta = 20$ fs at $\gamma_{\varepsilon_2}/\gamma_{\varepsilon_1} = 0.5$ (blue), $\gamma_{\varepsilon_2}/\gamma_{\varepsilon_1} = 1.0$ (orange), and $\gamma_{\varepsilon_2}/\gamma_{\varepsilon_1} = 1.5$ (green). Solid lines depict exponential fits according to Equation S3. The inset illustrates the spectral position of the ROI in the 2D spectra, as defined in Figure 3a. (b) Same traces as in (a) for the below diagonal cross peak ($\hbar\omega_\tau = 1.89$ eV, $\hbar\omega_t = 1.72$ eV). (c) Extracted fit time constant as a function of $\gamma_{\varepsilon_2}/\gamma_{\varepsilon_1}$ for the ROIs shown in (a) and (b). Semitransparent symbols represent values excluded from the corresponding fit due to pulse-overlap effects and electronic coherences.

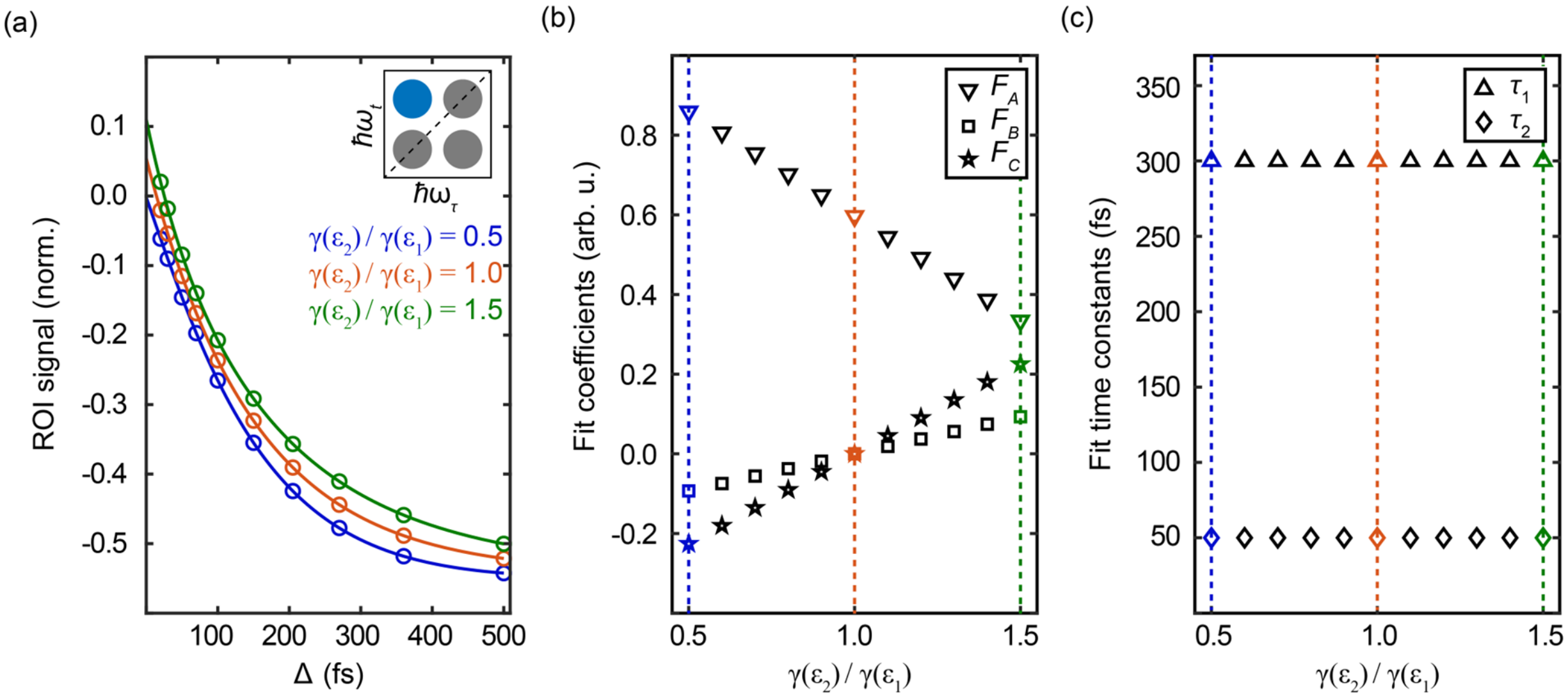

Figure S6. Dependency of the time-dependent cross peak ROI signal as a function of the delay time Δ on the ratio $\gamma_{\varepsilon_2}/\gamma_{\varepsilon_1}$. (a) Above-diagonal cross-peak ROI signal for $T = 20$ fs at $\gamma_{\varepsilon_2}/\gamma_{\varepsilon_1} = 0.5$ (blue), $\gamma_{\varepsilon_2}/\gamma_{\varepsilon_1} = 1.0$ (orange), and $\gamma_{\varepsilon_2}/\gamma_{\varepsilon_1} = 1.5$ (green), corresponding to the colored markers in (b) and (c). Solid lines depict exponential fits according to Equation S4, with $\tau_3$ fixed to the single-exciton energy transfer time (250 fs). The inset illustrates the spectral position of the ROI in the 2D spectra, as defined in Figure 3a. (b–c) Extracted fit amplitudes (b) and time constants (c) as a function of $\gamma_{\varepsilon_2}/\gamma_{\varepsilon_1}$.

A particularly interesting case is the time evolution of the high-energy diagonal peak (Figure S7). As discussed in the main text, ESA2-type pathways contribute weakly at early $T$, leading to weak EEA signatures. In our example, the single-exciton dynamics dominate the Δ-dependent time evolution at the spectral position of the high-energy diagonal peak for $\gamma_{\varepsilon_2}/\gamma_{\varepsilon_1} \geq 1.2$. Nevertheless, the fit results obtained using Equation S4 (Figure S7b and c) demonstrate that the time constants of the EEA kinetics can still be extracted. However, for a more reliable determination of the EEA dynamics, we recommend analyzing spectral regions with a pronounced EEA contribution, in particular the cross-peak positions at early $T$.

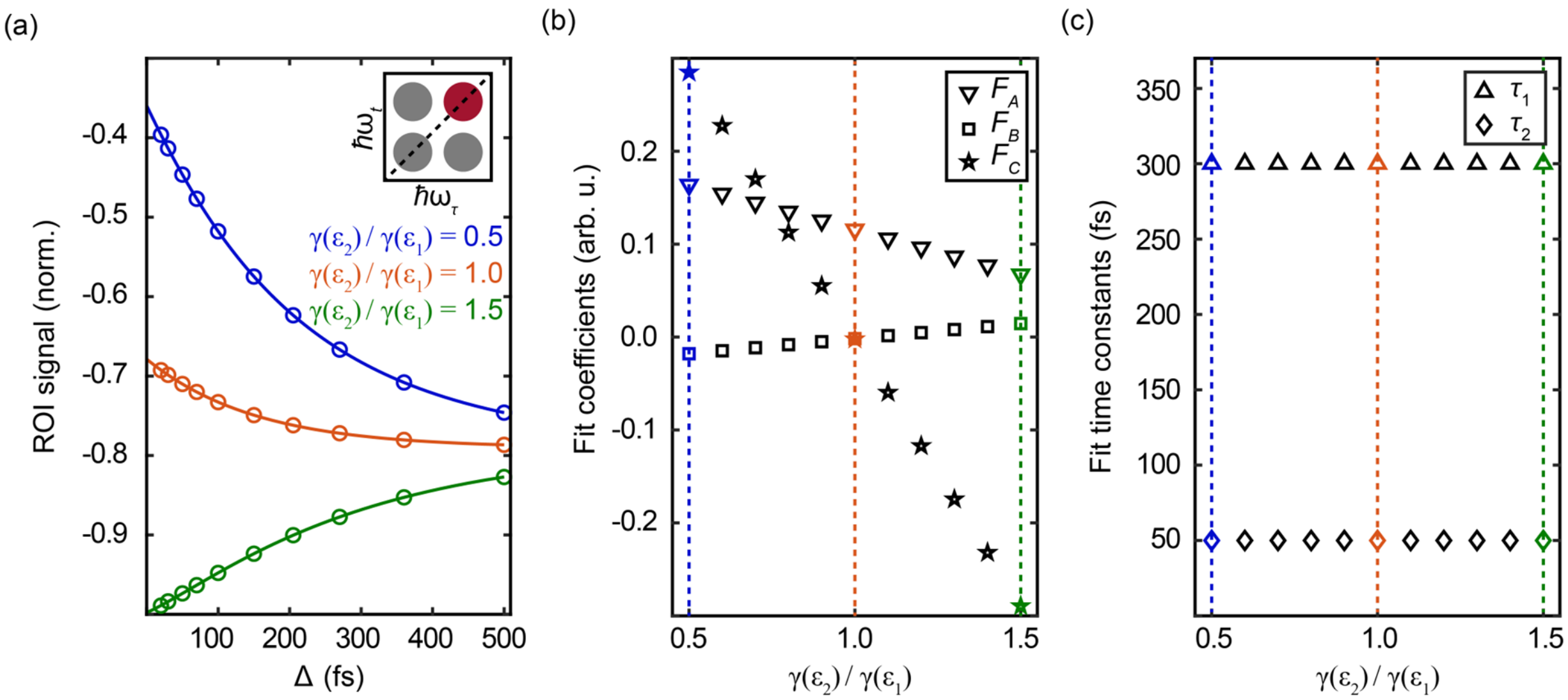

Figure S7. Dependency of the time-dependent high-energy diagonal-peak ROI signal as a function of the delay time $\Delta$ on the ratio $\gamma_{\varepsilon_2}/\gamma_{\varepsilon_1}$. (a) High-energy diagonal-peak ROI signal for $T = 20$ fs at $\gamma_{\varepsilon_2}/\gamma_{\varepsilon_1} = 0.5$ (blue), $\gamma_{\varepsilon_2}/\gamma_{\varepsilon_1} = 1.0$ (orange), and $\gamma_{\varepsilon_2}/\gamma_{\varepsilon_1} = 1.5$ (green), corresponding to the colored markers in (b) and (c). Solid lines depict exponential fits according to Equation S4, with $\tau_3$ fixed to the single-exciton energy transfer time (250 fs). The inset illustrates the spectral position of the ROI in the 2D spectra, as defined in Figure 3a. (b–c) Extracted fit amplitudes (b) and time constants (c) as a function of $\gamma_{\varepsilon_2}/\gamma_{\varepsilon_1}$.

## S7. Explicit comparison of C-2DES and F-2DES with P-2DES spectra

Figure S8 presents an explicit comparison of coherently detected, photoelectron-detected, and fluorescence-detected 2D spectra. For all spectra, the delay between the two pump pulses and the subsequent pulse was set to $T = 20$ fs. In Figure S8b, the delay of the ionization pulse was set to $\Delta = 20$ fs. As discussed in the main text, at early delay times $\Delta$ the photoelectron-detected 2D spectrum (Figure S8b) exhibits characteristic features of C-2DES (Figure S8a) with cross-peaks of alternating sign that indicate J-type coupling in this case. In contrast, the fluorescence-detected 2D spectrum (Figure S8c) shows cross-peaks with the same sign, arising from exciton–exciton annihilation (EEA) during the detection process. A more detailed comparison between coherently detected and fluorescence-detected 2DES is provided by Malý et al.[3]

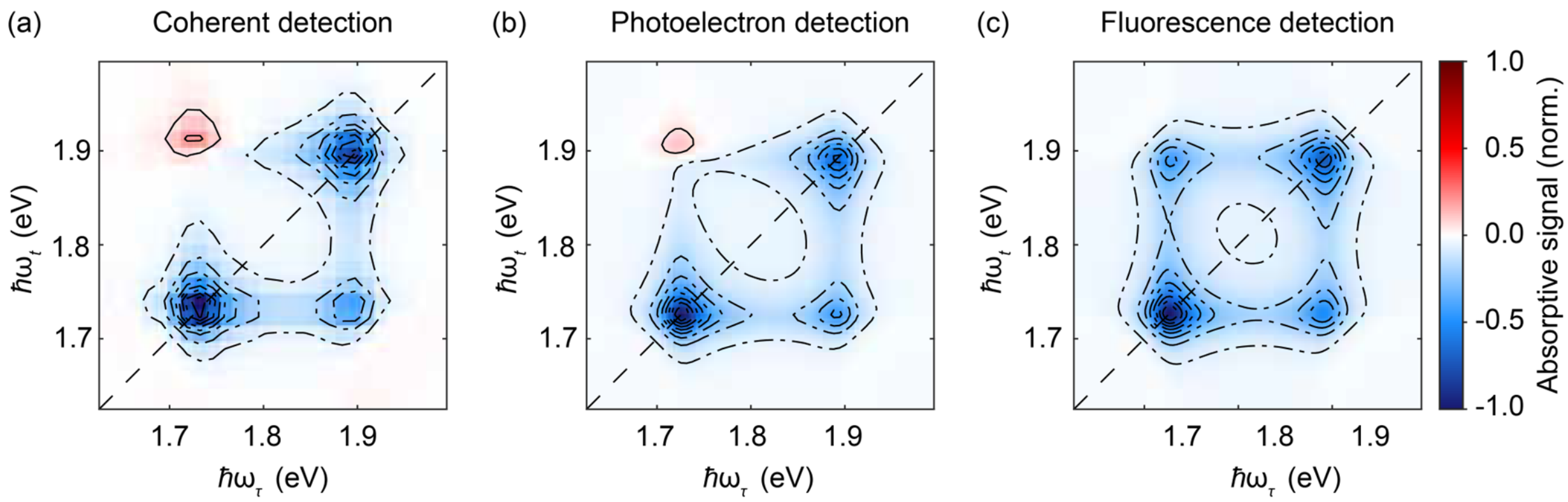

Figure S8. Comparison of 2D spectra obtained with different signal detection techniques. (a) Coherently detected 2D spectrum at $T = 20$ fs. (b) Photoelectron-detected 2D spectrum at $T = 20$ fs and $\Delta = 20$ fs. (c) Fluorescence-detected 2D spectrum at $T = 20$ fs.

## S8. Dependency of interexcitonic coherence signatures on the ionization pulse delay $\Delta$

During the delay time $T$, energy transfer as well as other exciton dynamics of the single-excited-state manifold can occur, such as, e.g., zero-quantum (0Q) coherences. These 0Q coherences include superpositions between various states with an energy gap smaller than or equal to the bandwidth of the excitation pulses, e.g., they represent vibronic or interexcitonic coherences. In our model system, the first two pulses generate a superposition of the two single-excited excitonic states $|\varepsilon_1\rangle$ and $|\varepsilon_2\rangle$, causing an oscillating feature for early delay times $T$. For better resolution of these coherences, we performed additional simulations for the early delay times $T$. Figure S9 shows the temporal evolution of the same ROIs as defined in Figure 3a. For better comparison, we use the value at $T = 150$ fs of each trace as a reference and subtract it from the corresponding trace. By comparison of the two traces of each subpanel, it becomes evident that the oscillating feature in each ROI is more pronounced for $\Delta = 500$ fs than for $\Delta = 20$ fs. Hence, the oscillating feature is more pronounced when efficient EEA occurs prior to the ionization pulse, which projects the final population state to the corresponding free-electron state. As described in the main text, destructive interference of the contributions SEc, ESAc, and ESA2c leads to a weakened coherence signature for early delay times Δ. However, at late delay times Δ, pathway cancellation of ESAc' and ESA2c' leads to a stronger contribution of the interexcitonic coherence to the total signal and, thus, to an enhanced contrast for these oscillating features. Bruschi et al. found a larger amplitude of coherent oscillations in F-2DES than in C-2DES.[7] In agreement with their observation, we find a higher sensitivity for large ionization pulse delays than for short ones.

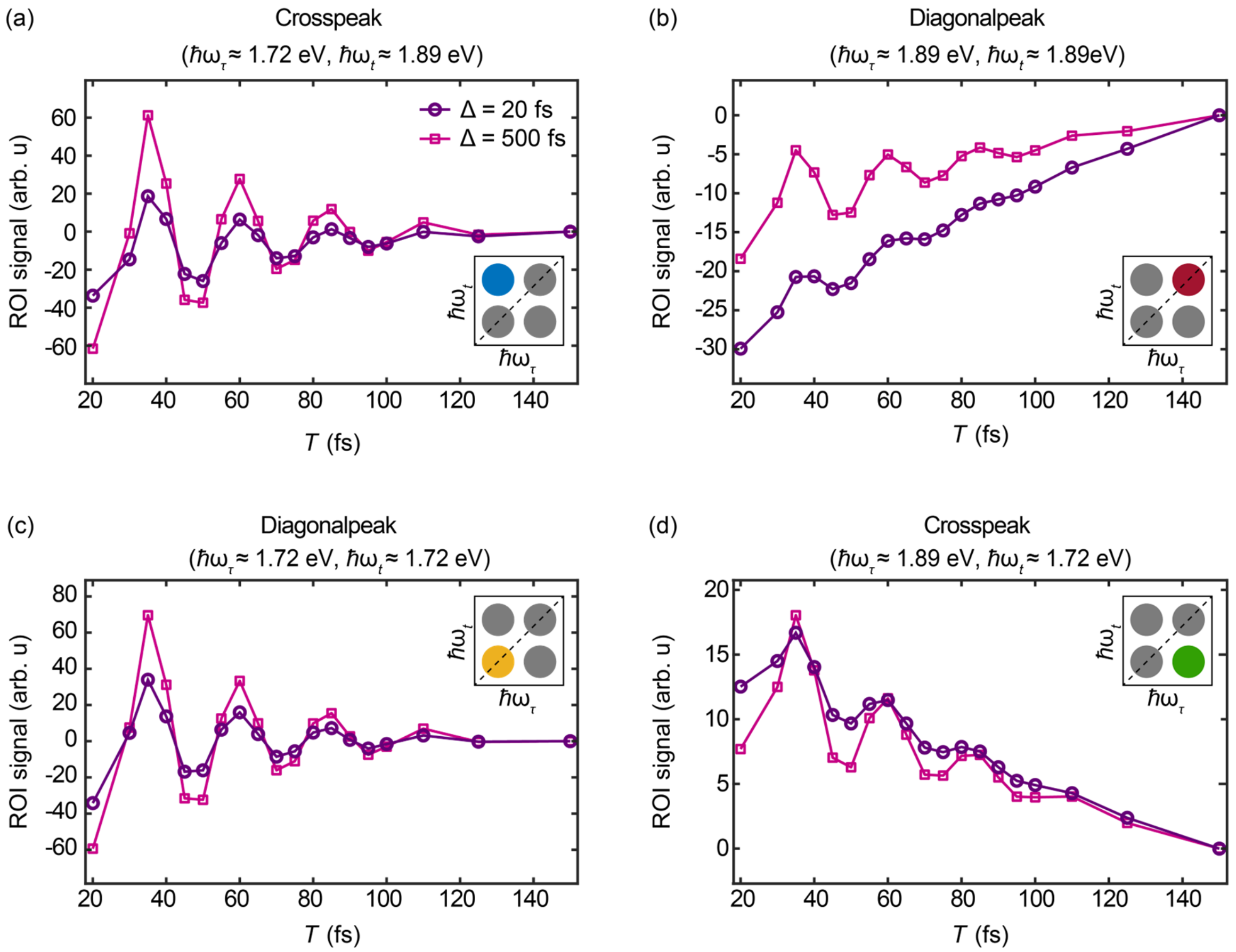

Figure S9. Early-time dynamics of the delay time *T*. Evolution of the signal integrated in the region of interest (ROI) as defined in Figure 3a for $\Delta = 20$ fs and $\Delta = 500$ fs. The value of each trace at $T = 150$ fs was used as reference point and subtracted from the the corresponding trace. Insets illustrate the spectral position of the ROI in the 2D spectra.

## S9. Dependency of annihilation signatures on energy transfer processes during $T$

In the main text, we discuss the contrast of the annihilation process during $\Delta$ for specific delay times $T$. Here, we exemplarily illustrate the sensitivity of the P-2DES signal against annihilation by comparing the signals of integrated ROIs for $T = 20$ fs and $T = 500$ fs (Figure 3c and S10, respectively) as function of the delay time $\Delta$. The individual ROIs are defined in Figure 3a in the main text. At both $T$ values, the above-diagonal cross-peak trace (blue circles) exhibits the smallest absolute amplitude but the strongest relative variation in signal strength between $\Delta = 20$ fs and $\Delta = 500$ fs, i.e., the strongest annihilation signature. This relative signal variation is similar for both $T$ values. In contrast, the signature of annihilation of the upper diagonal peak (red diamonds) strongly depends on the choice of $T$. The variation of the signal is only weak in comparison to the background for $T = 20$ fs. This background has nearly vanished at $T = 500$ fs and $\Delta = 20$ fs due to single-exciton energy transfer during $T$. With increasing delay time $\Delta$, the peak (re)emerges due to EEA, the variation in the signal is significantly stronger and has no static background. Consequently, the energy transfer during $T$ enhances the contrast of the annihilation signature of the above-diagonal peak.

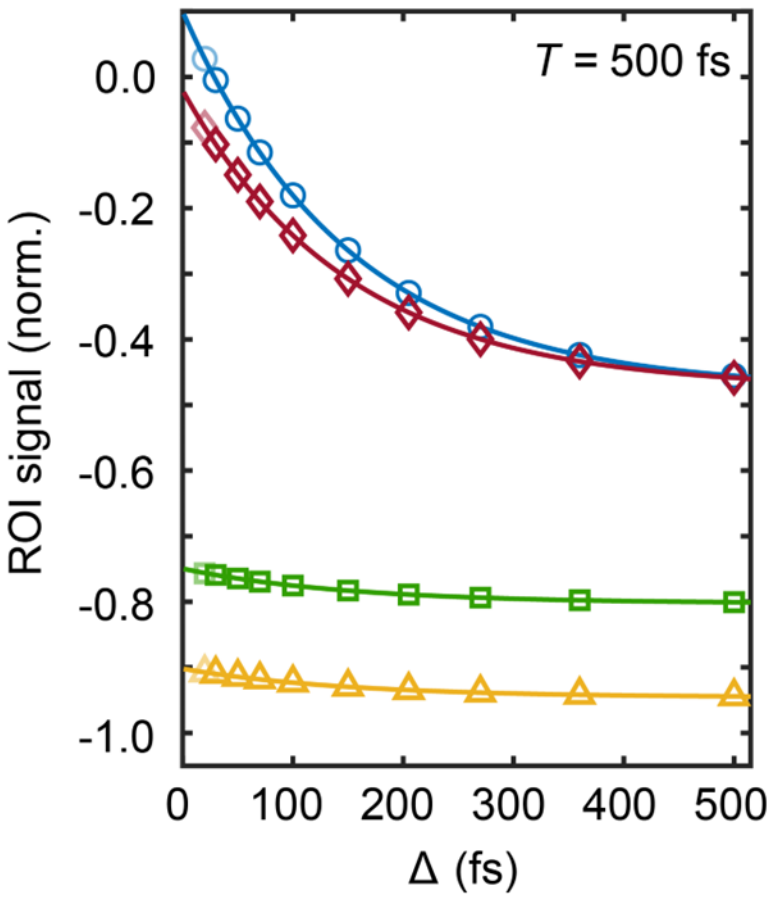

Figure S10. Time-dependent signal integrated over different ROIs for $T = 500$ fs as a function of the delay time $\Delta$. All ROIs are defined in Figure 3a. One ROI is drawn for the low- (yellow) and high-energy diagonal peaks (red) as well as for the cross peaks below (green) and above (blue) the diagonal. Solid lines depict exponential fits with a time constant of 150 fs. Semitransparent symbols represent values excluded from the corresponding fit due to pulse-overlap effects. All traces are normalized with respect to the maximum absolute value of all ROIs at $T = 20$ fs.

## S10. Rephasing Feynman diagrams of the high-energy diagonal peak

Figure S11 illustrates the rephasing Feynman diagrams of the high-energy diagonal peak. Apart from all diagrams that exhibit a coherence during $T$, each nonrephasing diagram (Figure 4) has a corresponding rephasing diagram. Notably, pathways exhibiting an interexcitonic coherence during $T$ contribute to the diagonal peaks in the rephasing signal, while the corresponding rephasing pathways contribute only at cross-peak positions.[8]

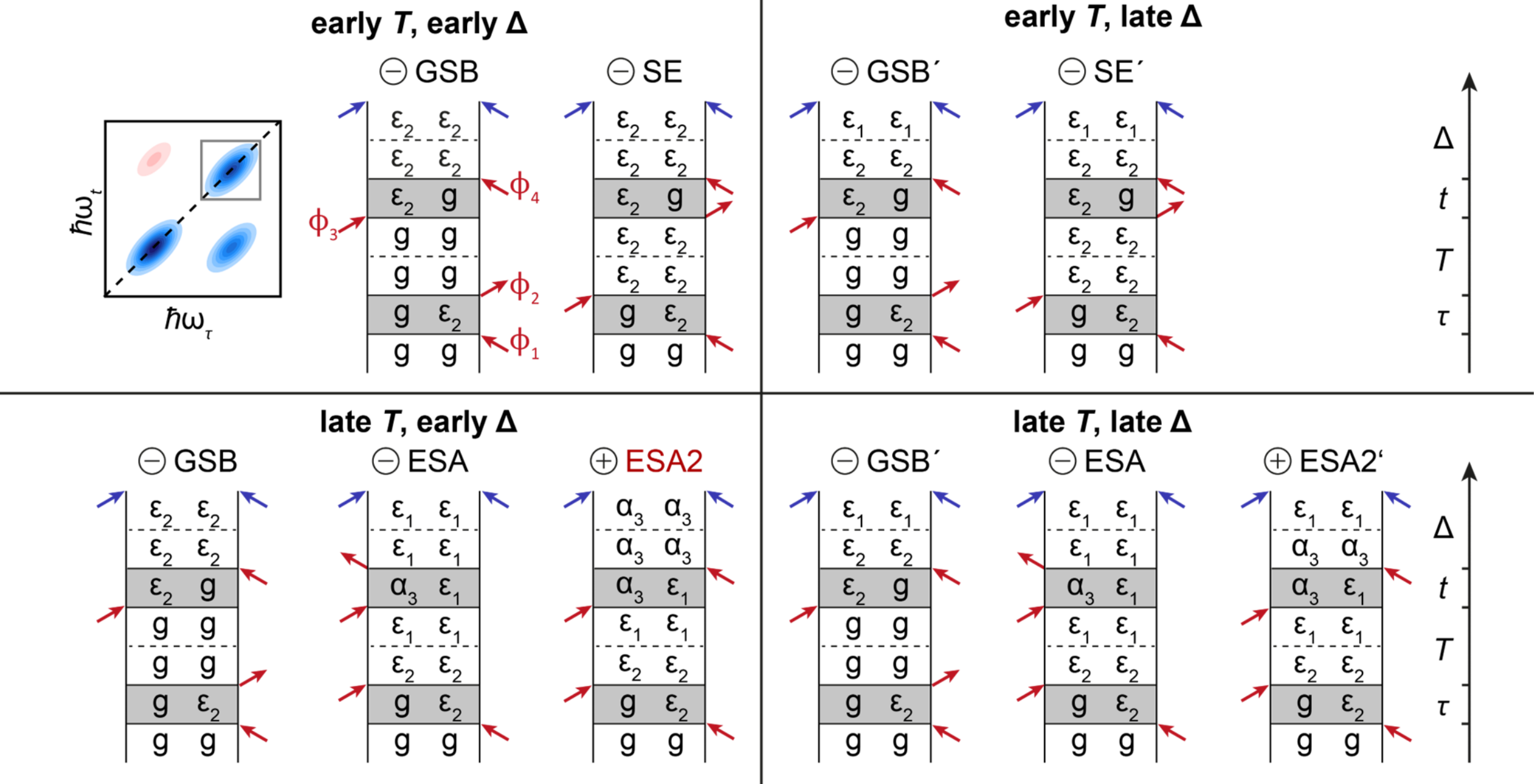

Figure S11. Rephasing Liouville pathways of the high-energy diagonal peak as depicted in the scheme on the top left corner. Four different cases are distinguished for the selection of delay times $T$ and Δ. Their groupings, separated by the vertical and horizontal black lines, correspond to the arrangement of data panels in Figure 3a. Time ordering of all pulses is assumed with respect to the sequence shown in Figure 1b of the main text. Red arrows represent interactions with the phase-cycled four-pulse sequence while blue arrows indicate interactions with the ionization pulse. Phase labels are only illustrated for the first diagram and are removed for all others for readability. Characters on the left-hand side of each represent ket state vectors; characters on the right-hand side represent the bra state vectors. The sign and type of each diagram are indicated above each diagram. If a pathway includes relaxation after the fourth pulse, they are labeled with a prime. All pathways contribute with $\Gamma_j = 1$, except the red-marked ESA2 pathways, which contribute with $\Gamma_{\alpha_3} = 2$.